**Title:**

# Physical limits of wind energy within the atmosphere and its use as renewable energy: From the theoretical basis to practical implications

**Short title:**

Physical limits of wind energy

**Authors:**

Axel Kleidon*

**Affiliation:**

Max-Planck-Institute for Biogeochemistry, Jena, Germany

*****Corresponding author:**

Axel Kleidon, Max-Planck-Institute for Biogeochemistry, Hans-Knöll-Str. 10, 07745 Jena, Germany, akleidon@bgc-jena.mpg.de

**Abstract**

How much wind energy does the atmosphere generate, and how much of it can at best be used as renewable energy? This review aims to give physically-based answers to both questions, providing first-order estimates and sensitivities that are consistent with those obtained from numerical simulation models. The first part describes how thermodynamics determines how much wind energy the atmosphere is physically capable of generating at large scales from the solar radiative forcing. The work done to generate and maintain large-scale atmospheric motion can be seen as the consequence of an atmospheric heat engine, which is driven by the difference in solar radiative heating between the tropics and the poles. The resulting motion transports heat, which depletes this differential solar heating and the associated, large-scale temperature difference, which drives this energy conversion in the first place. This interaction between the thermodynamic driver (temperature difference) and the resulting dynamics (heat transport) is critical for determining the maximum power that can be generated. It leads to a maximum in the global mean generation rate of kinetic energy of about 1.7 W m$^{-2}$ and matches rates inferred from observations of about 2.1 - 2.5 W m$^{-2}$ very well. This represents less than 1% of the total absorbed solar radiation that is converted into kinetic energy. Although it would seem that the atmosphere is extremely inefficient in generating motion, thermodynamics shows that the atmosphere works as hard as it can to generate the energy contained in the winds. The second part focuses on the limits of converting the kinetic energy of the atmosphere into renewable energy. Considering the momentum balance of the lower atmosphere shows that at large-scales, only a fraction of about 26% of the kinetic energy can at most be converted to renewable energy, consistent with insights from climate model simulations. This yields a typical resource potential in the order of 0.5 W m$^{-2}$ per surface area in the global mean. The apparent discrepancy with much higher yields of single wind turbines and small wind farms can be explained by the spatial scale of about 100 km at which kinetic energy near the surface is being dissipated and replenished. I close with a discussion of how these insights are compatible to established meteorological concepts, inform practical applications for wind resource estimations, and, more generally, how such physical concepts, particularly limits regarding energy conversion, can set the basis for doing climate science in a simple, analytical, and transparent way.

# 1. Introduction

In the current transition to a sustainable energy system, renewable forms of energy, such as solar, wind energy, hydropower, and biofuels, play a central role. Wind energy, the use of the kinetic energy associated with atmospheric motion by wind turbines, is one of the more common forms of renewable energy that is used today. It has seen a rapid expansion in the recent two decades. In Europe, for instance, the installed capacity of wind turbines has more than doubled over the last decade from 77 GW at the end of 2009 to 205 GW at the end of 2019 (WindEurope 2020). Some scenarios expect wind energy to continue to grow, considering 450 GW of installed capacity in offshore areas of Europe alone in 2050, with about half to be installed in the North Sea (WindEurope 2019). In Germany, wind energy on land has roughly doubled during the last decade, with an increase in installed capacity from 25.7 GW at the end of 2009 to 53.3 GW at the end of 2019, contributing more than 40% of the renewably generated electricity in Germany (BMWi 2020). Scenarios for 2050 envision the installed capacity of onshore wind energy in Germany to increase to 102 - 178 GW, with additional 51 GW - 60 GW installed offshore (BDI 2019, WWF 2019).

Such an anticipated increased use of wind energy in the future raises questions about the limits to wind energy use. How much can wind energy, at most, contribute to human energy needs? Can wind energy meet the entire energy needs of industrialized countries? Is wind energy so abundant that it can continuously power all human civilization, as some scientists have argued (Vance, 2009)? These are important questions which need to be answered. The basis for answering these relate to questions about how, and how much kinetic energy the atmosphere generates, and about the physical limitations of how much of the kinetic energy in the atmosphere can further be converted into renewable energy useful to human societies. These are climatological, rather than technical questions, and it would help to have simple, physically-based estimates as answers to these.

This review aims to provide a basis to answer these questions. While these questions have been addressed for a very long time (e.g., by Heinrich Hertz back in 1885 (as cited in Mulligan and Hertz 1997); by Lotka (1925), Lorenz (1955), and by Gustavson (1979)), what is different today is that we have far more observations, more sophisticated numerical simulation models, as well as advances in Earth system theory which allow us to identify the first-order controls and derive associated estimates to these questions. What I want to do here is to place these questions within a thermodynamic Earth system context (following Kleidon 2010, 2016). This approach focuses on the natural conversion of the energy contained in incoming solar radiation to the intermediate form of heat to kinetic energy of atmospheric motion and finally into renewable energy (as shown in Figure 1). These energy conversions are described and limited by physics, particularly by the second law of thermodynamics, but also by the conservation of energy, mass and momentum. The aim here is to provide quantitative descriptions of these conversions in a simple way, such that they can be used to derive the estimates given in Figure 1, and then link these to estimates derived from observations and sensitivities from numerical simulation models. In doing so, I want to demonstrate that first-order estimates can be derived from basic physics, and that these are consistent with recent studies with much more complex numerical simulation models.

This focus on first-order estimates and the large, planetary scale may at first appear to be at odds with the substantial heterogeneity in the wind fields, as observed over land, particularly in mountainous regions (see, e.g, the Global Wind Atlas, http://www.globalwindatlas.info), where, for instance, speed-up effects over mountain ridges shape wind speed patterns across the landscape. Such speed-up effects can be understood from Bernoulli's principle, an expression of the conservation of energy in the absence of friction. This principle states that when the flow speeds up due to an obstacle or a mountain range and its kinetic energy increases, this comes at the cost of a lower static pressure and potential energy. However, this change in energy is reversible, conserving the dry static energy of the flow (e.g., Ambaum, 2010). This means that the increase in wind speed does not represent a generation of kinetic energy from a diabatic heating source, but merely an adiabatic and reversible redistribution among the potential and kinetic energies within the flow. Such effects may thus be relevant for the spatial patterns of wind energy across heterogeneous landscapes, but they do not need to be taken into account when asking how much kinetic energy is generated by the atmosphere in the first place. When we look for the major mechanism that generates wind energy, this relates to the large-scale atmospheric circulation as this drives the synoptic-scale variations of high wind speeds in the mid-latitudes, and this, in turn, represents the dominant source for the wind energy currently being used as renewable energy. This wind energy is generated from large-scale differences in radiative heating due to insolation differences and represents the focus of this paper. It may not capture all sources of kinetic energy generation within the atmosphere, but it should capture the vast majority of it. With this focus on the planetary forcing, this review represents a top-down approach to estimating wind energy generation and its potential use as renewable energy.

The outline of this review is as follows. In section 2, I first focus on the conversion of solar radiation to kinetic energy to address the natural generation of kinetic energy in the atmosphere at the large scale. I will describe this conversion at a basic level, set up a simple box model to illustrate how thermodynamics and the Carnot limit yield a maximum kinetic energy generation rate associated with a certain value of poleward heat transport that is consistent with observations. I will then relate this description to relevant modelling studies, the proposed principle of Maximum Entropy Production (MEP), as well as the concept of the Lorenz energy cycle that describes the energy conversions associated with the maintenance of the atmospheric circulation.

This is followed by section 3, where I describe the physical limitations of the further conversion of kinetic energy from the atmosphere into renewable energy. I will start with the Lanchester-Betz limit, a well-known limit for the maximum efficiency of wind turbines, and extend it to a regional to planetary scale to make it consistent with the maximum generation of kinetic energy of the atmosphere derived in the first part. This extension is formulated in terms of another simple model which considers the limited supply rate of kinetic energy from the free atmosphere to the boundary layer, where this energy is either dissipated by friction or converted into renewable energy. This model yields estimates for the sensitivity of wind energy generation at large scales as well as its maximum. These are compared with estimates and sensitivities derived from climate models. At the end, I will then synthesize the two parts, aiming to provide the answers to the main questions of wind energy raised earlier, and conclude with a few implications.

## 2. What limits the generation of wind energy by the atmosphere at the large scale?

The typical starting point for describing atmospheric motion are the equations of fluid dynamics and the main determinants of these equations. Here, I want to take a step back, and focus on the energy source and the conversion sequence that generates kinetic energy to identify how thermodynamics limits this generation. As we will see, this provides a simpler, more holistic view of the factors that constrain atmospheric motion and its generation while remaining consistent with the Lorenz energy cycle (Lorenz 1955, Lorenz 1967). Most importantly, this approach will allow us to estimate how much kinetic energy the atmosphere generates in the climatological mean, solely based on physical concepts.

We start by looking at the source of kinetic energy: the radiative setting of the planet. Due to its geometry, the tropical regions of the Earth receive more solar radiation than the extratropics. This difference in solar radiative fluxes is well known and observed. Figure 2 (top) shows the climatological mean of satellite-based observations of the total absorbed solar radiation, which is the net flux of solar radiation at the top of the atmosphere (TOA). This data, represented by the red line, clearly illustrates the difference in solar radiative forcing across latitudes (data taken from the CERES dataset, Loeb et al, 2018; Kato et al., 2018).

When solar radiation is being absorbed, mostly at the surface, it is converted into thermal energy, or heat. This absorption takes place at the warm surface, while the emission from the atmosphere proceeds at a colder temperature. The resulting temperature difference is used to derive kinetic energy associated with dry and moist convection and to drive hydrologic cycling (Pauluis and Held, 2002), including losses by irreversible evaporation into an unsaturated atmosphere (Kleidon, 2008), dehumidification, and lifting of water before it is dissipated by falling raindrops (Pauluis et al., 2000). Because of the comparatively large energy fluxes in the vertical, in principle quite a bit of work can be derived from this forcing, although most of it is likely to be consumed by convection and the hydrologic cycle. In the following it is assumed that the power derived from the vertical differences do not contribute much to the maintenance of the large-scale, mostly horizontal circulation and rather focus on the horizontal differences in radiative heating, with a greater influx and absorption of solar radiation in the tropics compared to the extratropics. This generates a temperature difference between the tropics and the extratropics (Figure 2, bottom) that can also be used to derive power.

Where does the large-scale atmospheric circulation come into the picture? We can think of it as being a manifestation of the atmosphere transporting heat, aiming to level out the differences in solar radiative heating across the planet and the associated temperature difference. Kinetic energy needs to be generated against surface friction to accomplish this heat transport. This is where thermodynamics comes into play. We can think of the generation of kinetic energy as the consequence of the work done by a heat engine, a well-established thermodynamic concept that describes the conversion of heat into a form of free energy, defined here as the ability to perform work, such as the work associated with acceleration or lifting. This heat engine is driven by the difference in temperatures between the tropics and the extratropics and the heat flux into the heat engine is the heat transported by the atmospheric circulation.

This heat engine view of the atmosphere is schematically shown in Figure 3. As a consequence of the work performed by the engine, motion is generated and heat is transported from the tropics to the extratropics, aiming to level out the difference in radiative heating and to deplete the driving temperature difference. This transported heat is reflected in the outgoing flux of longwave (or terrestrial) radiation at the top of the atmosphere exceeding the absorbed solar radiation, a well-established sign for heat transport (cf. red and blue lines at the top of Figure 2). While absorption of solar radiation exceeds emission in the tropics, the opposite holds for the extratropics. The difference is the effect of heat redistribution by the climate system, primarily by atmospheric motion (and to a lesser extent by the oceans). It results in a more even distribution of the net emission of terrestrial radiation to space at the top of the atmosphere, and results in a more homogeneous distribution of surface temperatures (compare dotted line with dashed line at the bottom of Figure 2).

How much work can the atmospheric heat engine at best perform? Thermodynamics with its Carnot limit helps us to find the answer. In addition, however, we also need to account for the consequences of the heat engine, because the associated heat transport reduces the temperature difference and thereby the efficiency of the engine. For this, we need to include the energy balances to describe how the temperatures react to what the heat engine does and combine these with the Carnot limit. The combination of both then results in a maximum power limit (see, e.g., Kleidon, 2010), a limit we use here to infer how much kinetic energy the atmosphere can at best generate.

## 2.1 A simple energy balance model of the atmosphere

Our starting point is the formulation of the energy balances, or, more precisely, heat balances, to describe the surface temperatures of the tropics and extratropics, which we then combine with the Carnot limit. To do so, we use a simplified setup with two boxes to describe the Earth, represented by tropical and extratropical regions (Figure 3). I use the subscripts "*h*" for hot/tropics and "*c*" for cold/extratropics in the variables to describe the two boxes. The energy balances are formulated in their climatological mean state. The tropical region is shown by the red-shaded area in Figure 2 and roughly separates the latitudes with a net heat gain due to the net radiative flux at TOA (tropics) from those with a net heat loss (extratropics). The variables and the resulting estimates of kinetic energy generation are summarized in Table 1.

The net absorption of solar radiation, described by the fluxes $R_{s,h}$ and $R_{s,c}$, is the main forcing for the boxes. Averaging the CERES data shown in Figure 2 in the climatological mean over the respective latitudes yields values of $R_{s,h}$ = 306 W m$^{-2}$ and $R_{s,c}$ = 177 W m$^{-2}$ (see also Table 1). The cooling by emission is described by the fluxes of outgoing longwave radiation at TOA by $R_{l,h}$ and $R_{l,c}$. To parameterize these fluxes as a function of surface temperature, $T_s$, I use an empirical formulation in the form of $R_l(T_s) = a + b\, T_s$, with ($a$ = -388.7 W m$^{-2}$, $b$ = 2.17 W m$^{-2}$ K$^{-1}$) as in Budyko (1969). This empirical form has recently been shown to result from the linear scaling of the water vapor feedback with temperature (Koll and Cronin, 2018).

We can now set up the energy balances for each box to obtain expressions for the temperatures

$$R_{s,h} = R_{l,h}(T_h) + J \tag{2.1}$$

and

$$R_{s,c} + J = R_{l,c}(T_c) \tag{2.2}$$

where $J$ is the heat transport from the tropics to the extratropics. Using the empirical parameterization for $R_l(T_s)$, we can solve Eqns. 2.1 and 2.2 to obtain expressions for the temperatures:

$$T_h = \frac{R_{s,h} - a - J}{b} \tag{2.3}$$

and

$$T_c = \frac{R_{s,c} - a + J}{b} \tag{2.4}$$

Note how these equations reflect the effect of heat transport on the temperature difference, $T_h - T_c$: More heat transport (a greater value of $J$) depletes the temperature difference, because $J$ cools the tropics (a negative term in the expression for $T_h$) while warming the extratropics (a positive term in the expression for $T_c$).

We obtain the power generated by the atmospheric heat engine by using the Carnot limit with these expressions for the temperatures. The Carnot limit follows directly from the combination of the first and second law of thermodynamics. The first law describes the conservation of energy in the heat engine, as shown by the fluxes that enter and leave the rectangle marked "heat engine" in Figure 3. The heat input to the engine is represented by the heat taken from the tropics at a rate $J_{in}$, which is expelled from the engine in the extratropics at a rate $J_{out}$, with the difference being the power $G$ generated in form of kinetic energy. In steady state with no heat storage changes within the engine, the first law of thermodynamics then expresses the balance of these three rates:

$$J_{in} - J_{out} - G = 0 \tag{2.5}$$

Note the slight difference in notation for the heat engine, with $J_{in}$ and $J_{out}$, compared to the formulation of the energy balances, which simply uses $J$. This difference is resolved in steady state: The heat removed from the tropics represents $J$, as in Eq. 2.1, so that $J_{in} = J$, while the heat added to the extratropics in Eq. 2.2 is represented by $J_{out}$ by heat transport, but also by the frictional dissipation $D$, that is, the conversion of kinetic energy back into heat. This dissipation can be assumed to be in steady state, so that $G = D$. Then, the heat addition to the extratropics is also represented by $J = J_{out} + D = J_{in}$. Hence, this formulation of the heat engine is consistent with the energy balance formulation in Eqns. 2.1 and 2.2.

The other constraint on the heat engine is the second law, which is represented by the entropy budget of the heat engine. This budget is described by the entropy exchange across the boundaries of the heat engine, which add or remove heat with different entropy, and the entropy production, σ, within the heat engine. In our case, the entropy added to the heat engine is given by $J_{in}/T_h$, while the heat flux $J_{out}$ removes entropy at a rate $J_{out}/T_c$. The entropy budget of the heat engine is thus described by

$$\frac{dS}{dt} = \frac{J_{in}}{T_h} - \frac{J_{out}}{T_c} + \sigma = 0 \tag{2.6}$$

where *dS*/*dt* represents the change in entropy with time of the heat contained in the heat engine, which is assumed to average out to zero in the climatological mean. The second law demands that the entropy production within the heat engine cannot be negative, that is, σ ≥ 0. In the best case, no entropy production takes place within the heat engine, so σ = 0. This case represents the limit where most of the heat input can be converted into power. The entropy budget then simplifies to $J_{in}/T_h = J_{out}/T_c$, or $J_{out} = J_{in} T_c/T_h$. Combined with Eq. 2.5 and solved for *G* yields the well-known expression for the Carnot limit:

$$G = J_{in} \cdot \frac{T_h - T_c}{T_h} \qquad (2.7)$$

Note that in this derivation of the Carnot limit, no assumption was made about the actual thermodynamic cycle, as it is only based on the general consideration of the entropy budget of the system. Also note that this Carnot limit does not represent the limit of a dissipative heat engine, as described by Renno and Ingersoll (1996) and Bister and Emanuel (1998), for which the expression is slightly different by having a $T_c$ in the denominator, resulting in slightly more power. This limit emerges when the frictional dissipation is considered as an additional heating term at the warm side of the heat engine. In our case, however, frictional dissipation takes place mostly in the extratropics because of their higher wind speeds near the surface. Then, the frictional dissipation adds heating at the cold side of the heat engine (see text following Eq. 2.5), and does not yield extra power.

## 2.2 Maximum kinetic energy generation

To make the final step and infer the maximum in power that the atmosphere can generate at most, we combine the Carnot limit (Eq. 2.7) with the expressions of the temperatures (Eqns. 2.3 and 2.4). The heat flux $J_{in}$ corresponds to the *J* in the surface energy balance in Eqns. 2.1 and 2.3, although we need to reduce it by a factor of 1/2 for the global estimate, because the heat flux $J_{in}$ is obtained from the tropics, which represents half the surface area of the Earth.

With this, the power *G* has a clear maximum with respect to the magnitude of poleward heat transport, *J*. This is because as *J* gains in magnitude, the temperature difference between the tropics and the extratropics, $T_h$ - $T_c$, decreases, as reflected by the opposite signs in the energy balance equations (Eqns. 2.1 and 2.2) and the derived expressions for the temperatures (Eqns. 2.3 and 2.4). This sensitivity is shown in Figure 4. As a consequence, the efficiency term in the Carnot limit (the second factor on the right-hand side of Eq. 2.7 that involves the temperature difference) decreases with increasing *J*, resulting in the maximum in power.

Mathematically, this maximum in power is obtained by $dG/dJ = 0$, which can easily be derived when assuming that the temperature in the denominator of the efficiency term in Eq. 2.7 does not vary by much. It yields an optimum solution for the heat flux, $J_{opt}$,

$$J_{opt} \approx \frac{R_{s,h} - R_{s,c}}{4} \tag{2.8}$$

and the temperature difference, $T_{h,opt} - T_{c,opt}$, of

$$T_{h,opt} - T_{c,opt} \approx \frac{R_{s,h} - R_{s,c}}{2b} \tag{2.9}$$

Taken together, these yield a maximum in power, $G_{max}$, of

$$G_{max} = \frac{(R_{s,h} - R_{s,c})^2}{16 b T_h} \tag{2.10}$$

We use these expressions next to make quantitative estimates. For this, we need information on the difference in absorbed solar radiation, $R_{s,h} - R_{s,c}$ (for which we use the climatological mean of the CERES dataset, Fig. 2), a value of $b$ (from Budyko's (1969) parametrization), and a temperature (for which we use the mean surface temperature derived from surface emission in the CERES dataset). The resulting values are summarized in Table 1. They yield a maximum in power of about 1.6 W m$^{-2}$, or, when multiplied by the Earth's surface area of 511 x 10$^{12}$ m$^2$, a total of 800 x 10$^{12}$ W, or 800 TW. Using the CERES-derived heat flux, combined with the temperature differences inferred from surface emission for the Carnot expression given by Eq. 2.7, yields a similar value of about 850 x 10$^{12}$ W, or 850 TW, for the generation of kinetic energy.

Hence, only a small fraction of the incoming solar radiation can at best be converted into kinetic energy of the large-scale atmospheric circulation. We can express this fraction as the ratio of the maximum power to the amount of heat being transported and obtain an efficiency, $\eta$, in a way that is consistent with how efficiency is defined in thermodynamics as the energy generated in relation to the energy input. Using the expressions from above (2.8 and 2.10), this yields

$$\eta = \frac{G_{max}}{J_{opt}} = \frac{1}{2} \cdot \frac{T_{h,opt} - T_{c,opt}}{T_{h,opt}} = \frac{R_{s,h} - R_{s,c}}{4 b T_h} \tag{2.11}$$

with a value of about 5%. We note that it is only the difference in absorbed solar radiation that enters Eq. 2.11, as this difference relates directly to the difference in temperatures that drives the heat engine and shapes the efficiency term in the Carnot limit in Eq. 2.7.

When we relate the power to the total absorption of solar radiation, which is a much larger quantity, the efficiency is further reduced to 0.7%. This means that less than 1% of the absorbed solar radiation can at best be converted into kinetic energy associated with the large-scale atmospheric circulation. This low efficiency of conversion has long been recognized (e.g., Sverdrup, 1917; Ertel and Köhler, 1948). We should, however, note that this efficiency, while well established, is strictly speaking not a thermodynamic efficiency, because some of the reductions relate to the geometry of the planet that causes the differential radiative heating, and not directly to an inefficient conversion mechanism. In the following, I nevertheless use the term efficiency to also describe the fraction of the energy in relation to the total energy input by solar radiation.

## 2.3 Discussion

While this two-box model is a highly-simplified picture of the energetics of the atmospheric circulation, its overall behavior as well as the quantitative estimates are largely consistent with previously published research. Although the use of only two boxes may seem like a gross simplification, the basic insight regarding the maximum in power and the value associated with the maximum are essentially independent of the spatial resolution being used. This was already shown in previous research, where multi-box models (Paltridge 1975, 1978, 1979), two-box models (Lorenz et al. 2001, Kleidon 2004, Kleidon and Renner 2013a, b), or full numerical general circulation models of the atmosphere (Kleidon et al. 2003, 2006; Pascale et al. 2012) were used to demonstrate such outcomes. The underlying reason that the spatial resolution plays essentially no role is probably the relatively smooth, almost linear variation of the planetary forcing, so that the major thermodynamic forcing can even be captured by a minimum representation of two boxes and associated averages of the forcing.

On the theoretical side, the mechanism resulting in maximum power is essentially the same as the one which results in a state of maximum entropy production (MEP), a proposed principle for the organization of complex, non-equilibrium thermodynamical systems (e.g., Dewar, 2003; Ozawa et al., 2003; Kleidon and Lorenz, 2005; Martyushev and Seleznev, 2006; Kleidon et al. 2010), which has been applied to the atmosphere for several decades (starting with Paltridge, 1975, 1978, 1979). The link between our maximum power estimate and the maximum in entropy production is made by considering that in steady state, power balances frictional dissipation ($G = D$), and frictional dissipation adds heat to the atmosphere, which in turn produces entropy. This rate of entropy production is given by (using Eq. 2.7)

$$\sigma = \frac{D}{T_c} = \frac{G}{T_c} = \frac{J_{in}}{2} \cdot \frac{T_h - T_c}{T_h T_c} \qquad (2.12)$$

Starting with Paltridge (1975), a series of studies over the last decades used the maximization of entropy production to maximize an expression similar to Eq. (2.12) to infer atmospheric heat transport and derive the climatological state on Earth (see, e.g., review by Ozawa et al. 2003), or other planetary bodies (Lorenz et al. 2001). While energy balance models are typically used, such maxima can also be found in general circulation models of the atmosphere, for instance by varying semiempirical parameters related to friction (Kleidon et al., 2003, 2006; Pascale et al., 2012). The maximization of power, as done here, is almost identical to the maximization of entropy production, with the difference being the influence of the $1/T_c$ term in Eq. 2.12 compared to Eq. 2.7. The estimate of kinetic energy generation from such a thermodynamic approach is thus essentially the same, regardless of whether maximum power or maximum entropy production is used. The picture of the atmosphere acting as a heat engine that maximizes the power to generate kinetic energy seems, however, easier to comprehend than an abstract maximization of entropy production.

The maximum in power can be related to the well-established framework of the Lorenz Energy Cycle (LEC, Lorenz 1955, 1967), a meteorological concept that describes the generation of kinetic energy from available potential energy at large scales that dates back to Margules (1905). The LEC framework, however, is typically not related to the broader context, specifically the generation of available potential energy from the diabatic heating source of differential radiative heating. This link is established by considering the potential energy of an air column in hydrostatic equilibrium, which is proportional to its internal energy (or heat) content, with the proportionality given by the gas constant of air, $R_a = 287$ J kg$^{-1}$ K$^{-1}$, in relation to the heat capacity at constant pressure, $c_p$. The difference in temperatures between the tropics and the extratropics thus translates directly into a difference in potential energy.

However, not all of this potential energy difference is available for conversion into kinetic energy. Typically, estimates of available potential energy are based on adiabatic considerations. Our thermodynamic approach allows for a somewhat different, yet consistent alternative by recognizing that only a fraction of $(T_h - T_c)/T_h$ represents heat that can be converted into kinetic energy. With this fraction, we can then infer a magnitude of available potential energy associated with the two-box model above. Using the same surface pressure $p_s$ for both boxes ($p_s = 1013.25$ hPa), the available potential energy, $A$, associated with maximum power translates into

$$A = \frac{1}{2} \cdot R_a \frac{p_s}{g} \cdot (T_{h,opt} - T_{c,opt}) \cdot \frac{T_{h,opt} - T_{c,opt}}{T_{h,opt}} \tag{2.13}$$

with $R_a$ = 287 J kg$^{-1}$ K$^{-1}$ being the gas constant for air, $p_s/g$ representing the atmospheric mass per unit area and $g$ = 9.81 m s$^{-1}$ being the gravitational acceleration. The factor of 1/2 is, again, included to account for area weighting.

Using the values from Table 1, this yields a numerical value of about 46 x 10$^5$ J m$^{-2}$. This magnitude is largely consistent with estimates of available potential energy of 42 - 45 x 10$^5$ J m$^{-2}$, which have been derived from observations and large-scale atmospheric reanalysis products (e.g., Oort 1964; Li et al. 2007; Boer and Lambert 2008). One should note, however, that the available potential energy is not fixed, but depleted by the heat transport (as is the temperature difference). Lorenz (1960) already noticed this aspect when he explored the maximum in the generation rate of available potential energy, although not in the context of an atmospheric heat engine maximizing its power output.

The estimate of kinetic energy generation derived from maximum power also compares well to observation-based estimates of the Lorenz Energy Cycle and the associated kinetic energy generation rate. Recent estimates based on reanalysis data place this rate at around 2.1 - 2.5 W m$^{-2}$ (Li et al. 2007, Boer and Lambert 2008), consistent with a thermodynamically-based estimate of this magnitude that dates back to Lettau (1954). This estimate is close to the estimate of 1.6 W m$^{-2}$ derived here, albeit slightly higher. Also, the magnitude of heat transport of 32 W m$^{-2}$ determined from maximum power is less than the inferred magnitude from the CERES data (as shown by the vertical grey and blue-shaded areas in Figure 4, also Table 1), which yields a value of 44 W m$^{-2}$ for the average imbalance of the net fluxes of solar and terrestrial radiation at the top of the atmosphere. Consequently, the temperature difference $T_{h,opt}$ - $T_{c,opt}$ ≈ 30 K is larger than the mean difference inferred from the CERES dataset of $T_h$ - $T_c$ ≈ 21 K.

This low bias in our maximum power estimate may have several reasons. Of course, the extremely simple formulation of the model here and the use of the climatological mean forcing are strong simplifications of the system. To start, annual means were used to infer the maximum in power, rather than seasonal, or monthly averages. This may yield an underestimate because in the winter season, the differences in absorbed solar radiation as well as temperatures are both enhanced, so that this covariation among the two terms in the Carnot limit (Eq. 2.7) does not average out. However, since the seasonal variations of solar radiation in the extratropics are strongly buffered by heat storage changes in the surface ocean, the temperature difference $T_h - T_c$ does not respond as strongly to variations in radiative forcing and heat transport as in the absence of these heat storage variations. This reduced sensitivity by the buffering effect of heat storage changes can be approximated by using the annual mean temperature differences. Then, only the first term in the Carnot limit shows a seasonal variation as does the power, and it can be averaged over the seasons. Hence, the annual means used here appear to be adequate in estimating the mean kinetic energy generation of the atmosphere even though the flux and temperature differences appear to show a covariation. A full treatment of heat storage changes and seasonal variations could, of course, be explicitly included in a numerical formulation, but this would make the estimates necessarily more complex and less transparent.

Also, no contribution by the Hadley circulation to generate kinetic energy is accounted for. The Hadley circulation generates its power from the vertical temperature difference between the surface and the radiative temperature of the atmosphere, as described at the beginning of this section, rather than from the horizontal temperature difference. Since the motion associated with the Hadley cell is associated with kinetic energy, but driven from a convective heat engine that is primarily driven by condensational heating, this component would not be included in the estimate derived here. Likewise, tropical cyclones generate high wind speeds near the surface out of the condensational heating of moist air (e.g., Emanuel 1999). Both aspects are driven by heating from condensation of atmospheric moisture, which in turn is maintained by surface evaporation. As this energy derives directly from surface absorption of solar radiation (rather than differences), this would contribute some additional kinetic energy at the large scale. The free energy generated by moist convection is also constrained by a maximum power limit (see e.g., Kleidon and Renner, 2013a), and is mostly associated with driving the hydrologic cycle, as described earlier. Further work would be needed to estimate how much the Hadley circulation contributes to the generation of kinetic energy at large scales.

Regional circulations, such as land-sea breeze circulations, are also not included in our estimate. When comparing the estimate derived from maximum power to the much more detailed estimate of kinetic energy generation based on the LEC framework (e.g., Li et al. 2007), these additional processes may contribute additional kinetic energy generation of up to 60% (($2.5$ W m$^{-2}$ – $1.6$ W m$^{-2}$) / ($1.6$ W m$^{-2}$)) to the overall generation of the atmosphere. This, however, does not seem to drastically alter the magnitude derived here from the highly simplified 2-box model, so that these factors should not affect the implications that can be drawn from this strong thermodynamic limitation.

Furthermore, dynamic constraints, particularly related to the conservation of angular momentum, are not included as well, a point raised previously as a critique to the MEP hypothesis (Goody 2007). Angular momentum conservation, which, with its Coriolis force, causes the mid-latitudinal belt of westerly winds, would seem to act as a possible further constraint that could be included in the maximization of power (see Jupp and Cox (2010); Pascale et al. (2013) for works that include the effect of rotation in the application of MEP to the climate system). On the other hand, the heat transport by the ocean has not been accounted for, although it contributes to the overall heat transport towards the poles and which would lessen the power generated in the atmosphere. These factors add a certain level of uncertainty to the estimate derived here.

## 2.4 Summary

In summary, the combination of a simple energy balance model with the thermodynamic Carnot limit yields a first-order estimate of how much kinetic energy the atmosphere can at best generate in the climatological mean. This estimate of about 1.6 W m$^{-2}$, or 800 TW globally, is consistent in magnitude with the well-established estimates of 2.1 - 2.5 W m$^{-2}$, so a magnitude of 2 W m$^{-2}$ is used in the summary shown in Figure 1. It would thus appear that the atmosphere operates close to its thermodynamic limit, working as hard as it can to generate kinetic energy.

It represents a physical picture that can explain the low conversion efficiency of solar radiation into the kinetic energy of atmospheric motion. The low overall efficiency is set by the difference in solar absorption being the primary driver, rather than the mean absorption of solar radiation, in combination with the low temperature difference that yields a low thermodynamic conversion efficiency. The Carnot efficiency of the conversion is reduced further by the interaction of the resulting atmospheric motion that transports heat and depletes the driving temperature difference that is created by the uneven absorption of solar radiation.

What this interpretation emphasizes is the importance of a systems approach that includes the basic interactions to evaluate how much kinetic energy the atmosphere can generate at best. The resulting low conversion efficiency of 0.7% from absorbed solar radiation into kinetic energy is thus not a reflection of an inefficient atmosphere, but rather the manifestation of an atmosphere that works as hard as possible to generate the most kinetic energy, given the low temperature differences available for the conversion, and which is further depleted by the vast amounts of heat that the atmosphere transports.

# 3. What are the limits of using wind energy as renewable energy at the large scale?

We have now linked the well-established magnitude of about 2 W m$^{-2}$ of kinetic energy generation of the atmosphere in the global climatological mean to the thermodynamic limit of maximum power, implying that this is the most wind energy that the atmosphere can generate from the radiative forcing of the planet. We next turn to the second part of this paper, dealing with the limits of the further conversion into renewable energy.

Before we get to this limit, I want to first motivate the importance of characterizing this limit, because there are recently published resource potentials for wind energy available that are above the mean kinetic energy generation rate of the atmosphere, so that one may question the physical assumptions used in these resource assessments. For this, I use two examples of such resource assessments to illustrate this discrepancy.

The typical methodology in estimating regional wind energy potentials is based on populating a region with individual turbines with a certain spacing, use the turbine characteristics and combine them with wind fields to estimate yields. From the atmospheric side, the main contribution for such estimates are the prevailing wind fields, and these wind fields are assumed to remain unaffected by how many turbines are being considered. A recent study on the wind energy resource for Europe (Enevoldsen et al. 2019), for instance, used such an approach to conclude that turbines with a total of 52.5 TW of installed capacity distributed over half of Europe (4.9 x $10^6$ km$^2$) could yield 1.4 x $10^5$ TWh of electricity per year. This corresponds to a mean electricity generation rate of (1.4 x $10^5$ TWh/a) / (8760 h/a) / (4.9 x $10^6$ km$^2$) = 3.3 W m$^{-2}$. Similarly for Germany, the German environmental protection agency ("Umweltbundesamt", UBA) in 2013 used such an approach (UBA, 2013), selected suitable areas, and estimated that 2900 TWh of electricity could be generated per year from turbines with a cumulative installed capacity of 1190 GW distributed over an area of 49 400 km$^2$ (representing about 14% of the area of Germany). This estimate would correspond to a mean electricity generation rate of (2900 TWh/a) / (8760 h/a) / (49.4 x $10^3$ km$^2$) = 6.7 W m$^{-2}$. Both examples represent cases in which turbines would, on average, yield more electricity per unit area than the ≈ 2 W m$^{-2}$ of kinetic energy that the atmospheric circulation generates and dissipates in the climatological mean. We may thus ask how the atmosphere responds when wind energy is used at such scales, what the impact of this is on the yield of large assemblages of wind turbines, rather than a single turbine or a small wind farm, and how it affects the estimate for the resource potential of a region.

Our starting point is the wind energy generation of a single turbine (Figure 5, shown by the box labeled "Turbine limit"). Wind with a certain speed $v$ and air density $\rho$ represents a flux density of kinetic energy, given by the expression

$$J_{ke} = \frac{1}{2} \rho v^3 \tag{3.1}$$

with units of W m$^{-2}$ through a cross-section perpendicular to the flow. This flux density is sometimes referred to as wind power density, although here I prefer the more accurate term of kinetic energy flux density as it does not deal with power, defined as work performed per time, but rather to a flow of energy.

The rotor of a wind turbine represents a cross-sectional area, $A_{rotor}$, to this flow, withdraws a certain fraction of the contained kinetic energy, and converts it into electricity at a certain rate $G_{turb}$. The limit of how much of the kinetic energy flux can at best be converted into renewable energy has been well researched and answered. The Lanchester-Betz limit of a wind turbine (Lanchester 1915, Betz 1920, see also van Kuik, 2007) describes that only a fraction of about 59% can at best be taken out from the flow and converted by the turbine into electricity. This limit is derived from the mass and momentum balance constraints of the flow. A brief summary of its derivation is provided in Appendix A. These constraints tell us that the conversion efficiency $\eta_{turb}$ from wind to renewable energy of a single turbine is at best given by (using the expression equivalent to the efficiency of kinetic energy generation, Eq. 2.11):

$$\eta_{turb} = \frac{G_{turb}}{J_{ke}} \leq 59\% \tag{3.2}$$

Current turbines, when operating above their cut-in velocity but below their rated capacity, typically have an efficiency $\eta_{turb}$, or, power coefficient, of about 40-50% (Carrillo et al., 2013; also Germer and Kleidon, 2019, Fig. S3), which is slightly lower than the Lanchester-Betz limit.

The Lanchester-Betz limit, however, only deals with the limit of a single turbine — it does not consider the factors that maintain the kinetic energy flux from which the turbines draw their energy, nor does it consider the effect of the removal of energy by the turbines on this flow, thereby leaving an impact on the atmosphere (Gans et al. 2012). The inclusion of these processes in a broader system's view of turbines and atmosphere is critical when we consider more and more turbines, up to the scale of regional wind energy resource estimates.

This removal effect by wind turbines and their effect on wind energy resource potentials at larger scales has clearly been shown with climate model simulations in which the effect of many turbines on the atmosphere is explicitly simulated. This has been done at the global scale (Miller et al., 2011; Jacobsen and Archer, 2012; Marvel et al., 2012; Miller and Kleidon, 2016), at the regional scale (Keith et al., 2004, Adams and Keith 2013, Miller et al. 2015), or for wind farms of different sizes, ranging from 25 to $10^5$ km$^2$ in scale (Volker et al., 2017). The electricity generation that these simulations yield are consistent with atmospheric energetics, because they explicitly resolve the dynamics that generate and transport kinetic energy within the atmosphere (as described in the first part above) and include the atmospheric response to wind energy use. These simulations consistently show that while the electricity generation rate increases with more intense wind energy use either by employing more turbines within an area or by extending the area being considered, they also show that the mean generation rate per unit area converges against a limit of about 1 W m$^{-2}$ or less. This was, for instance, shown by model simulations for the central US by Adams and Keith (2013) and Miller et al (2015), or at the global scale by Miller and Kleidon (2016). There may be exceptions in some regions, e.g., the very windy southern tip of South America (Volker et al. 2017), yet a large-scale limit of less than 1 W m$^{-2}$ for wind energy use is consistent with atmospheric energetics in that it is less than the 2 W m$^{-2}$ that the atmosphere generates in the mean. The key atmospheric effect that explains this limit at larger scales are wind speed reductions due to the larger number of wind turbines that are present within the region (Miller and Kleidon, 2016).

How can we account for such wind speed reductions in a simple, physical way, and obtain an energetically-consistent estimate for how much wind energy can be used as renewable energy at large-scales? In the following, we set up another simple model that focuses on how the kinetic energy generated by the atmosphere is either dissipated in the boundary layer due to friction or is converted into renewable energy by turbines. Dissipation by friction depends on wind speed, so the more kinetic energy is diverted to renewable energy, the more frictional dissipation is reduced and so is the wind speed near the surface. We use this approach to derive the limit of the conversion into renewable energy, with the formulation being described in Miller et al. (2011), Gans et al. (2012), Miller et al. (2015), and Miller and Kleidon (2016). It results in a limit that takes a broader system's perspective than the limit of a single turbine, as illustrated in Fig. 5 by the box labelled "Atmospheric limit". We will see in the following that this approach leads to quite a lower limit on how much wind energy can at best be generated than typical wind energy resource assessments, yet yields estimates that are consistent with climate model simulations and atmospheric energetics.

## 3.1 A simple momentum balance model of the lower atmosphere

We start with a minimum formulation of the momentum balance as a basis to describe how kinetic energy is being converted in the lower atmosphere (the VKE model described in Miller et al. (2011), Gans et al. (2012), Miller et al. (2015), and Miller and Kleidon (2016), but extended here to account for all dissipative losses in the boundary layer). We consider the climatological mean state, so that temporal changes in momentum average out, and consider a region sufficiently large such that net horizontal advection of momentum is negligible. Then, the balance is given by

$$\int_{z_{bl}} \frac{d(\rho v)}{dt} dz = 0 = F_{down} - \tau - F_{turb} \qquad (3.3)$$

with the left-hand side being the temporal change in horizontal momentum averaged over the boundary layer height $z_{bl}$ (which is assumed to average out to zero in the climatological mean), $F_{down}$ being the downward transport of horizontal momentum from the free atmosphere, $\tau$ is the surface stress due to friction, and $F_{turb}$ is the total force exerted by all wind turbines to remove kinetic energy from the flow.

The surface stress $\tau$ can be expressed by the commonly used drag formulation in form of

$$\tau = \rho C_d v^2 \tag{3.4}$$

where $\rho$ is the air density, $C_d$ is the drag coefficient, and $v$ is the wind speed in the lower atmosphere. By combining Eqns. (3.3) and (3.4), we obtain an expression for the wind speed of

$$v = \sqrt{\frac{F_{down} - F_{turb}}{\rho C_d}} \tag{3.5}$$

In this expression, we can immediately see the effect of more wind turbines (a larger value of $F_{turb}$) in reducing wind speeds.

What happens with the kinetic energy supplied from the free atmosphere in this formulation? When horizontal momentum is transported downwards, starting from an initially higher wind speed $v_{free}$ of the free atmosphere, it supplies kinetic energy to the near-surface atmosphere at a rate

$$J_{v,ke} = F_{down} \cdot v_{free} \tag{3.6}$$

with $J_{v,ke}$ being the vertical flux of kinetic energy per surface area. The kinetic energy is mixed with the flow of the boundary layer, resulting in dissipation by downward mixing, $D_{mix}$, given by

$$D_{mix} = F_{down} \cdot (v_{free} - v) \tag{3.7}$$

The remaining kinetic energy is then either dissipated by surface stress, $D_{surf}$,

$$D_{surf} = \tau \cdot v \tag{3.8}$$

or it is taken out by the wind turbines to generate electricity, $G_{turb}$, and the associated dissipation in the wake, $D_{wake}$,

$$G_{turb} + D_{wake} = F_{turb} \cdot v \tag{3.9}$$

The budgeting of kinetic energy in the lower atmosphere then demands that these rates balance:

$$J_{v,ke} = D_{mix} + D_{surf} + D_{wake} + G_{turb} \tag{3.10}$$

In the absence of wind turbines, $D_{wake}$ and $G_{turb}$ are zero, so that all dissipation occurs either between the free atmosphere (with wind speed $v_{free}$) and the wind speed $v$ at a reference height within the boundary layer (dissipation $D_{mix}$), or between the reference height and the surface (dissipation $D_{surf}$).

Note that $G_{turb}$ refers here to the cumulative yield of turbines per surface area, and not to the yield of an individual turbine (as in Appendix A).

## 3.2 Maximum energy yield by wind turbines

We can now use this momentum balance formulation to determine the maximum of kinetic energy that can at best be used by wind turbines. To do so, we maximize $G_{turb}$ with respect to $F_{turb}$ by setting $dG_{turb}/dF_{turb} = 0$. This maximization is done with a prescribed downward flux of horizontal momentum $F_{down}$ from the free atmosphere. The existence of a maximum results from the trade-off between a greater generation rate of the turbines due to a greater value of $F_{turb}$ (Eq. 3.9), but a reduced wind speed (Eq. 3.5). When we assume that wake dissipation of the turbines is 50% of the generated electricity, $G_{turb}$, the proportion associated with the Lanchester-Betz limit (see Appendix, Eq. A13), all we need for the maximization is the information of the wind speed $v$ in the absence of wind turbines as well as the downward flux of horizontal momentum $F_{down}$ (note that it is this additional information on $F_{down}$ which makes this approach different to the common resource potential estimation described in the beginning). The maximization can be done analytically, and yields an optimum, total force exerted by the turbines per unit area of

$$F_{turb,opt} = \frac{2}{3} \cdot F_{down} \qquad (3.11)$$

At this maximum, the velocity is reduced to

$$v_{opt} = \frac{1}{\sqrt{3}} \cdot v_0 \qquad (3.12)$$

which is about 58% of the wind speed $v_0$ with no wind turbines present. The maximum generated power is then given by

$$G_{turb,max} = \frac{4}{9\sqrt{3}} \cdot F_{down} \cdot v_0 \qquad (3.13)$$

which represents 26% of the dissipation near the surface $D_{surf}$ with no turbines present.

We can reformulate this limit in terms of a conversion efficiency as before, by dividing the generated power of the turbines by the energy supply, which is the kinetic energy that would naturally be dissipated near the surface in the absence of wind turbines:

$$\eta_{turb} = \frac{G_{turb}}{D_{surf}} \approx 26\% \tag{3.14}$$

Note that this efficiency is not quite comparable to the Lanchester-Betz limit. While the Lanchester-Betz limit describes the maximum power that can be taken out by a single turbine out of an undisturbed flow, the limit given by Eq. 3.13 accounts for the competition between the energy extraction by wind turbines and the natural frictional dissipation near the surface. Because of the inclusion of this competition, this limit does not just require the wind speed, but also the vertical momentum momentum flux $F_{down}$. This momentum flux can be estimated from $\tau$, because these two fluxes need to be equal due to momentum conservation in the absence of wind turbines. Both limits have nevertheless in common that they are entirely based on conservation laws and maximization, and do not require specific information of the turbine characteristics.

## 3.3 Quantifying wind energy limits

To quantify large-scale wind energy limits using the approach just described and test it against climate model simulations, we next use climatological fields of the ERA-5 reanalysis (Copernicus Climate Change Service, 2017) as inputs and then compare these to the simulations by Miller and Kleidon (2016). The goal in doing so is not just to get a global estimate for the wind energy resource potential, but also to show that the relatively simple formulation in the previous part yields sensitivities consistent with numerical model simulations. Furthermore, it provides insights as to why wind energy potentials drop below 1 W m$^{-2}$ per unit surface area when more and more wind turbines extract kinetic energy from the atmosphere at larger and larger scales.

The climatological fields used in the following evaluation are summarized in Figure 6, with the resulting estimate provided in Table 2. Shown are the mean wind speed $v$ (at 10m height, Fig. 6a), with winds being substantially higher over the ocean than over land (with median speeds of $v_{ocean}$ = 7.0 m s$^{-1}$ and $v_{land}$ = 3.9 m s$^{-1}$), the associated kinetic energy flux densities (with medians of $J_{ke,ocean}$ = 455 W m$^{-2}$ and $J_{ke,land}$ = 77 W m$^{-2}$ per square meter of cross-sectional area, Fig. 6b), the drag coefficients (with medians $C_{d,ocean}$ = 0.0013 and $C_{d,land}$ = 0.0065, Fig. 6c), and the overall frictional dissipation within the boundary layer (Fig. 6d), which represents the downward transport of kinetic energy from the free atmosphere to the surface (with medians of $J_{v,ke,ocean}$ = 1.8 W m$^{-2}$ and $J_{v,ke,land}$ = 2.1 W m$^{-2}$). Note how similar the frictional dissipation is for ocean and land in their respective means, and that both values are essentially the same magnitude as the generation rate of kinetic energy derived in the section above. The difference in wind speeds and kinetic energy flux densities thus arises mostly due to the difference in drag coefficients, which reflects the difference in aerodynamic roughness of the surfaces.

We next use the median values for these variables (as summarized in Table 2) to obtain the limit for wind energy use. Instead of using the medians of the wind speeds directly, I use an inferred wind speed from the medians of the kinetic energy flux densities, as these are more representative of the magnitude of surface friction, being directly related by $D_{surf} = 2\, C_d\, J_{ke}$ (cf. Eqns. 3.1, 3.4, 3.8). Using this approach yields median wind speeds of $v_{ocean}$ = 9.4 m s$^{-1}$ and $v_{land}$ = 5.2 m s$^{-1}$ (calculated with an air density of $\rho$ = 1.1 kg m$^{-3}$), which are notably higher than the median wind speeds from the climatological mean.

With the median drag coefficients and wind speeds being specified, we can determine the limit to wind energy use (as given by Eqns. 3.11 – 3.14), but also more generally infer how wind speeds (Eq. 3.5) and the generated power (Eq. 3.9) should vary with the number of wind turbines. We compare this sensitivity against the simulations by Miller and Kleidon (2016). They simulated this sensitivity with climate model simulations for a range of installed capacity densities of wind turbines, using the properties of $G_{cap}$ = 2 MW wind turbines with a rotor-swept area of $A_{rotor}$ = 2827 m$^2$ and assuming a Lanchester-Betz efficiency of $\eta$ = 59%. These wind turbines add a drag to the lowest atmospheric model layer. The total force exerted by the turbines is described by

$$F_{turb} = N \cdot \eta_{turb}\, A_{rotor} \cdot \frac{\rho}{2} v^2 \qquad (27)$$

where $N$ is the turbine number density (in units of turbines per m$^2$). The installed capacity density is then described by $N\, G_{cap}$ (in units of MW km$^{-2}$ or W m$^{-2}$).

The impact of increasing the turbine density on wind speed and wind energy generation is shown in Figure 7. The analysis is performed separately for land and ocean due to the differences in the drag coefficient and wind characteristics. The comparison shows that the VKE approach, driven with the ERA-5 global median values, can capture the global mean response of the much more detailed climate model simulations of Miller and Kleidon (2016) rather well, in terms of the reduction of wind speeds with greater wind energy use (specified by a greater installed capacity in the figure), and with a greater impact on the resulting yields.

We also note a considerable deviation between the simple estimate and the climate model simulations over oceans, particularly regarding the estimated electricity generation (Figure 7b). This discrepancy can be explained by different intensities in vertical mixing over ocean and land caused by the difference in how the diurnal variation of solar radiative input is being buffered (Kleidon and Renner, 2017). Oceans absorb solar radiation below the surface because water is transparent, causing heat storage changes in the surface ocean and little warming of the surface during the day. Land, however, absorbs solar radiation at the surface, heats up, generates buoyancy and boundary layer mixing, so that the buffering takes place mostly by heat storage changes in the lower atmosphere rather than below the surface. This difference in mechanism manifests itself in a much stronger variation of turbulent heat fluxes and air temperature during the day (Kleidon and Renner, 2017). The relevance of this difference is that over the ocean, the kinetic energy of the free atmosphere is transported downwards mostly diffusively, i.e., by vertical gradients in wind speed. Over land, the downward transport is dominated by the strong vertical mixing associated with the convective boundary layer during the day. When wind energy is intensively used, then a consequence of this is that the vertical downward transport is enhanced over the ocean because the reduced wind speeds caused by wind turbines enhances the velocity difference and hence the downward transport. On land, the kinetic energy is transported convectively, and is therefore mostly unaffected. This interpretation of an increased downward momentum transport over ocean was shown in the simulations in Miller and Kleidon (2016), while a similar effect can also be found over land during night when convective transport is absent (Miller et al., 2015).

In the next step, I want to link this analysis back to the question of how much wind energy can be used at large scales and how this question is linked to the limit of kinetic energy generation described in the first part. The comparison in Figure 7b shows that the simple estimate captures the limit of wind energy generation in the climate model simulations quite well, with an estimated maximum yield of 0.30 and 0.26 W m$^{-2}$ over ocean and land respectively (with additional 0.15 and 0.13 W m$^{-2}$ dissipated by wakes). These generation rates represent merely 17% and 12% of the kinetic energy input from the free atmosphere (cf. Table 2). This is lower than the 26% limit derived above.

To understand this discrepancy, we look at the different ways by which the kinetic energy input from the free atmosphere is either dissipated or used as renewable energy. The input of kinetic energy from the free atmosphere, $J_{v,ke}$, as diagnosed from the ERA-5 reanalysis (boundary layer dissipation, shown in Figure 6d), is shown in Figure 7c,d by the horizontal blue lines labeled "Total". This input is assumed to be fixed in the following analysis, implying that the consequences of wind turbines remain confined to the boundary layer and that the ability of the atmosphere to generate kinetic energy is not affected.

The kinetic energy input from the free atmosphere is either dissipated by boundary layer mixing, surface friction, wake dissipation, or converted into renewable energy by the wind turbines, as described by Eqns. 3.7 – 3.9. In the absence of wind energy use and with low installed capacities, the kinetic energy input is mostly dissipated by bringing it down into the boundary layer (purple lines in Figure 7c,d) and by surface friction (red lines). The proportion between these two forms of dissipation is shifted towards greater surface dissipation over the ocean, which can be attributed to the smoother surfaces with their lower drag coefficients than for land. In the presence of wind turbines, some of this energy is not dissipated by these two terms, but converted into renewable energy. With an increasing number of wind turbines, a greater proportion of the kinetic energy input is converted into renewable energy (black lines), with some loss by wake turbulence behind the turbines (orange lines). As some kinetic energy is removed from the atmosphere surrounding the turbines, wind speeds are reduced, and so is surface friction (red lines). With the reduced wind speeds near the surface, the dissipation by mixing above the turbines (purple lines) increases because of the greater wind speed differences. As a consequence, a considerable and increasing fraction of the kinetic energy input will be dissipated above the wind turbines. Hence, the maximum rate of kinetic energy that is diverted to renewable energy is relatively low with about 0.3 W m$^{-2}$ per surface area, representing roughly 12-17% of the kinetic energy input (Table 2). At this maximum, wind speeds decline by 42% (= 1 - 3$^{-1/2}$), as predicted by the VKE model (Eq. 3.12), which is consistent with the range of about 40-50% simulated by atmospheric general circulation models (Miller and Kleidon 2016, Table 1).

This illustration and the associated estimates were done using the ERA-5 wind speeds at 10m height. Current wind turbines, however, reach to greater heights, with typical hub heights of modern wind turbines being about 100m. At these levels, wind speeds are higher, less of the kinetic energy input is dissipated above the turbines, so that the estimate done above is likely too low. Yet, the sensitivity just described nevertheless remains qualitatively the same when the hub heights of wind turbines are shifted to greater heights, although the balance of the terms is somewhat altered. At the limit, 26% (cf. Eq. 3.14) of the kinetic energy input from the free atmosphere can be converted into renewable energy. While this conversion limit can yield quite different potentials across regions, depending on the level of dissipation (as shown in Figure 6d), we can use this limit also to evaluate the mean resource potential at the planetary scale. With a global mean generation rate of kinetic energy of about 2 W m$^{-2}$, this then yields an upper, large-scale limit of ≈ 0.5 W m$^{-2}$ that can at best be converted from wind into renewable energy (as shown in Figure 1).

## 3.4 Discussion

Naturally, the VKE model described above is very simple to estimate how much wind energy can at best be used. Despite some deficiencies, the comparison in Figure 7 shows that it captures the main mechanism of much more complex climate model simulations rather well. Yet, it also adds on to the first part of this paper, because the kinetic energy generated within the free atmosphere at large scales serves as the input for this estimate. In this view, local sources for generating circulations, such as sea or mountain breezes, were neglected, although these can also provide sources for wind energy in some regions.

Leaving this aspect of local circulations aside, what I want to do in the following is to first link the wind energy estimates to the energetics of the atmosphere to provide a plausible interpretation for the low estimated potentials when moving to larger scales. I then discuss two related aspects: the role of mid- to high altitude winds as a potential source of wind energy (using airborne wind energy technology), and the scale at which wind energy potentials decline from the high yields observed at small wind farms to the much lower, large-scale potentials.

The first part showed how the solar forcing and thermodynamics set the limit of generating kinetic energy within the atmosphere at large scales to about 2 W m$^{-2}$. This energy forms the input to the lower atmosphere, where it is naturally dissipated by friction, and converted back into heat. This physical picture is consistent with the notion that the large-scale dynamics of the free atmosphere is geostrophic, that is, the assumption that friction can be neglected.

The generated kinetic energy of about 2 W m$^{-2}$ then forms the energy input to the lower atmosphere. The conversion into renewable energy competes with the frictional dissipation, either near the surface (the red lines in Fig. 7c,d), or above the turbines (the purple lines in Fig. 7c,d). These unavoidable dissipative processes within the atmosphere can explain why the large-scale wind energy potential estimated by atmospheric models are substantially lower, with a range of 0.3 to 0.6 W m$^{-2}$, compared to large-scale estimates that use prescribed, observed wind fields, which yield wind resource estimates in the order of 4.3 to 6.2 W m$^{-2}$ (ranges taken from Table 1 in Miller and Kleidon, 2016).

What this tells us is that the low wind energy potentials at the large scale are not set by the particular technology that is being used, but rather by atmospheric limitations set by physical principles: the low ability of the atmosphere to generate kinetic energy from the differential solar radiative forcing, which is the consequence of the thermodynamic limit of maximum power, combined with the limited ability of converting kinetic energy to renewable energy in the presence of dissipative, frictional losses in the lower atmosphere, which is a consequence of the momentum balance. What this shows is that it requires a system's perspective to evaluate wind energy potentials at larger scales that includes the drivers and conversions of kinetic energy, from the solar energy input to its removal from the atmosphere by the wind turbines.

We can extend this system's perspective to address the question of wind energy availability at greater heights in the middle to upper atmosphere. Airborne wind energy generation aims to exploit this energy, with one of the main motivations being the presumably higher potential, justified by the higher wind speeds (e.g., http://airbornewindeurope.org/, Archer and Caldeira, 2009; Bechtle et al. 2019). The case in which we consider heights that still reside within the boundary layer can be linked to the above discussion of the VKE sensitivity shown in Figure 7c,d. With greater heights within the boundary layer, wind speeds would typically increase, and with this greater height, less of the kinetic energy is dissipated above and more is dissipated below this height (this would shift the magnitudes of the red and purple lines shown in Fig. 7c,d). This would lead to a somewhat greater potential for wind energy, yet still being constrained to the 26% of the energy input from the free atmosphere, as has already been described above.

When wind energy generation takes place in the free atmosphere, then this has quite different effects. It is likely to impact atmospheric dynamics substantially as it effectively adds a friction term to the geostrophic balance of the free atmosphere. This was, for instance, shown in climate model simulations by Miller et al. (2011b), which evaluated the potential and atmospheric effects of wind energy generation in the jet streams. By placing a kinetic energy sink into the jet stream, the simulations showed that this sink would cause an ageostrophic component to the flow, causing poleward heat transport in the upper atmosphere and thereby substantially altering the atmospheric circulation. These alterations reduced the kinetic energy generation by the atmosphere by about 1/3, while the potential to generate renewable energy would be very small, with on average less than 0.1 W m$^{-2}$ of generated renewable energy. Such strong impacts on atmospheric dynamics were also shown in simulations by Marvel et al. (2012), although one should note that the energetics in their simulations deviated substantially from observation-based estimates, with kinetic energy generation about 50% higher and mean kinetic energy being a quarter of what is estimated from observations. What this illustrates is that higher wind speeds at greater heights, particularly in the jet streams, are not associated with a greater wind energy potential, can cause substantial changes to atmospheric dynamics, and likely reduce the ability of the atmosphere to generate kinetic energy. Surface-based use of wind energy can also affect climate when used at large-scales (e.g., Keith et al. 2004, Miller et al. 2011), but it would seem that the magnitude and scale of the impact is much less. This can be attributed to the fact that the kinetic energy would be dissipated near the surface anyway, so its diversion to wind turbines would cause relatively little impact, except for affecting the regional wind field.

How does this low estimated wind energy potential fit together with the observed, high yields of single turbines and small wind farms?  Small wind farms in prominent locations can yield more than 6 W m$^{-2}$ per surface area (MacKay, 2013). As the scale of wind farms increases, the observed yield typically drops to values below 0.5 W m$^{-2}$ per surface area for scales of 100 km$^2$ or larger, as shown by the observational analysis from the US by Miller and Keith (2018).  This general trend towards lower yields per surface area has also been shown in model simulations (e.g., Volker et al. 2017), although a few regions may show higher yields of up to 3.5 W m$^{-2}$ per surface area because they are notably windier (like the southern tip of South America).  However, at the global scale, such very windy conditions are quite rare.  Boundary layer dissipation, which reflects the kinetic energy input from the free atmosphere, is less than 4 W m$^{-2}$ for 81% (87%) of the ocean (land) grid cells, which can be inferred from the frequency distribution shown in Fig. 6d.  Of this input, only a fraction can be used at large scales, so that a large-scale yield of 3.5 W m$^{-2}$, as reported by Volker et al. (2017) represents an exception, but is not representative for most parts of the planet.

The decrease in mean turbine yield per unit area at increasing spatial scales can be explained by considering the kinetic energy balance of the boundary layer that includes a net horizontal advection of kinetic energy (Kleidon and Miller, 2020), an approach that reproduced the yield reductions reported in Volker et al. (2017) very well. To describe this balance, we first draw a virtual box, as depicted in Figure 8, around the wind turbines that are being considered, with a certain width $W$, downwind length $L$, and the height of the box being the height of the boundary layer $H$.  We then consider the kinetic energy supply by the air flow through the horizontal cross section ($\rho/2\ v^3$ x $W$ x $H$, with $\rho/2\ v^3$ being the kinetic energy flux density, Eq. 3.1) and the supply from the free atmosphere, which would be dissipated by surface friction in the absence of turbines (hence given by $\rho\ C_d\ v^3$ x $W$ x $L$, with $\rho\ C_d\ v^3$ being the dissipation by surface friction, Eqns. 3.4 and 3.8, and which is much smaller as indicated in Fig. 8).  This kinetic energy supply is then balanced by the energy the turbines remove, losses by surface friction, mixing, and wake dissipation, and the downwind export of kinetic energy.

When we next consider the effect of wind farm size, particularly with respect to the downwind length $L$, then we note that the kinetic energy supply by the horizontal flow remains the same, because it does not depend on $L$, and that only the much smaller supply from the free atmosphere increases. The kinetic energy budget for a small wind farm with a small $L$ is then dominated by the large, kinetic energy supply through the vertical cross section ($W$ x $H$), which is consistent with the typical approach to wind resource estimation that is based on the horizontal kinetic energy flux density.  The importance of this horizontal supply diminishes as we consider larger wind farms with greater $L$, for which the small, vertical supply becomes more important, and hence the mean velocity of the flow as well as turbine yields need to decrease.

The relative importance of these two kinetic energy supply rates is described by the ratio $H/(2\ C_d\ L)$. For a single turbine and small wind farms, this ratio is large because of a low $L$, indicative of the horizontal flux supplying much more kinetic energy than the vertical flux. At successively larger downwind lengths, this ratio decreases, and the supply of kinetic energy from above becomes increasingly important. The large-scale limit, as described by the VKE model above, then represents the case where $L \to \infty$, so that there is no net supply of kinetic energy from upwind areas.

This, then, explains why the mean yield per unit area declines with larger wind farms with a greater downwind length $L$. While the kinetic energy supply rate through the upwind cross section is large (a median value of $J_{ke}$ = 455 W m$^{-2}$ per cross-sectional area over ocean, Table 2, Fig. 6b), the renewal from above is much smaller ($J_{v,ke}$ = 1.8 W m$^{-2}$, Table 2, Fig. 6d). When the box gets larger by increasing the downwind length $L$, the high yields cannot be maintained, but must decline. This is because the kinetic energy input does not increase at the same pace by which the surface area of the box increases. Consequently, the reservoir of kinetic energy within the boundary layer is depleted, resulting in lower wind speeds, which in turn cause lower mean yields of the turbines. When the downwind length gets very large, the kinetic energy input is dominated by $J_{v,ke}$, so that at large scales, the yield converges to the large-scale limit described by the VKE model above.

From the ratio of kinetic energy supply rates we can also derive a length scale, $L_d = H/(2\ C_d)$, at which kinetic energy is supplied at the same rates by the horizontal and vertical inputs (Kleidon and Miller, 2020). With a typical magnitude of boundary layer heights of $H \approx 1$ km and a typical drag coefficient of about $C_d \approx 0.005$, this yields a length scale of $L_d \approx 100$ km. This length scale characterizes the decline of high yields of small wind farms to the lower values of $\approx 0.5$ W m$^{-2}$ of wind energy potential at large scales that was derived here.

### 3.5 Summary

In summary, what I showed here is that the momentum balance of the lower atmosphere combined with a fixed kinetic energy generation rate sets a large-scale limit to wind energy use of about 0.5 W m$^{-2}$ in the global mean. In contrast to the well-established Lanchester-Betz limit, this limit is set by the competing effects of frictional dissipation and energy extraction by wind turbines spread over a large region that takes place within the atmospheric boundary layer. Energetic considerations define this scale to be around 100 km of wind energy use, with yields diminishing already before this scale of wind energy use is reached.

The estimate described here extends and complements the limit to wind energy generation described in the first part above. It shows that it does not require complex simulation models to capture the main effects that set the low wind resource potentials at large scales, as shown by the agreement between the estimates derived here from the momentum balance and those obtained with much more complex climate model simulations. What it emphasizes though is that it requires a system's perspective to estimate the wind energy resource at large scales, which accounts for the interaction with the atmosphere and the limited ability of the atmosphere to generate and transport kinetic energy. This can be done by high-resolution numerical modelling of atmospheric motion interacting with wind farms, but also by simply considering basic physical principles.

## 4. Summary and Conclusions

In this paper, I addressed the questions of how much wind energy the global atmosphere generates, and how much of this could be used as renewable energy. To provide answers, I used two simple, physics-based models that describe the conversion processes from differential solar absorption to kinetic energy (the first part), and the further conversion from kinetic energy to frictional heating and renewable energy (the second part). Both cases represent a system's approach, in the sense that the effects of the conversions on the overall setting were included.

In the first part, the generation of kinetic energy was estimated using a thermodynamic approach which views the atmosphere as a heat engine, fueled by differential solar radiative heating between the tropics and the poles. The system's approach enters here by including the main effect of the energy conversion into kinetic energy, the poleward heat transport associated with atmospheric motion, into the estimate. This heat transport depletes the large-scale temperature difference between the tropics and the extratropics, and therefore the thermodynamic driver of the conversion process. It reduces the efficiency term in the Carnot limit, and results in a maximum power limit of generating kinetic energy. This limit yields a global mean estimate of 1.6 W m$^{-2}$ of kinetic energy generation, which is of similar magnitude as the well-established estimates of 2.1 - 2.5 W m$^{-2}$ based on observations and reanalyses. Furthermore, this limit is associated with a certain, optimum magnitude of poleward heat transport and an intermediate equator-pole temperature difference, both of which are roughly consistent with observations. This first-order derivation of the magnitude of kinetic energy generation suggests that the atmosphere works as hard as it can to generate motion, and can explain the low resulting conversion efficiency from solar radiation into motion. For practical purposes, it establishes the fact that the atmosphere cannot generate additional kinetic energy for the current solar radiative forcing of the planet if a substantial fraction of wind energy is used as renewable energy.

The second part dealt with the further conversion of kinetic energy into renewable energy. Here, the system's perspective enters when the limited supply rate of kinetic energy is combined with dissipative losses and the conversion into renewable energy. The unavoidable effect of more and more wind turbines is that the stock of kinetic energy in the lower atmosphere is depleted, which is associated with lower wind speeds. When this effect is included in the estimation of how much wind energy can at best be converted into renewable energy, it results in a further limit of 26% of the supplied kinetic energy, resulting in a global mean estimate for the wind energy potential of about 0.5 W m$^{-2}$ at large scales (as summarized in Figure 1). This large-scale limit is approached when wind energy is intensively used at length scales greater than 100 km.

The resulting large-scale potential for wind energy merely represents 0.2% of the 240 W m$^{-2}$ of solar radiation that is on average absorbed by the Earth system. I want to briefly put this wind energy potential into perspective by comparing the total conversion efficiency from the energy input by solar radiation to electricity with two other forms of renewable energy (Figure 9). As shown in the first part, a major factor that results in this low efficiency is associated with the conversion of solar radiation into atmospheric motion. This low efficiency results from the fact that most of the low entropy of solar radiation, which is associated with the high emission temperature of the Sun, is lost when it is absorbed and converted into heat at Earth's prevailing surface temperatures. Then, it is the combination of the difference in solar heating (rather than the total) in combination with a comparatively low temperature difference, which is further depleted by atmospheric heat transport, that results in the low conversion efficiency of 0.7% from solar radiation to wind energy. The further conversion into renewable energy by wind turbines competes with frictional dissipation, resulting in a maximum conversion efficiency of 26% at large scales. This, in total, yields a global wind energy potential with an overall very low conversion efficiency of 0.2%.

Bioenergy is another form of renewable energy, which relies on photosynthesis. Photosynthesis converts solar radiation into chemical energy, rather than heat, so the conversion process is very different to atmospheric motion. While its efficiency can be quite high with 34% under some conditions (Hill and Rich, 1983), the efficiency of carbon fixation by terrestrial photosynthesis is low, with a median value of 0.8% for terrestrial ecosystems estimated from observations (Kleidon, 2021) of which only about half is further converted into biomass (with the other half consumed by plants by autotrophic respiration). The resulting biomass could then be converted further into electricity via combustion. Using an optimistic conversion efficiency of 60% would result in an overall conversion efficiency of about 0.2%, which is also rather low.

A sharp contrast is photovoltaics, where observations from the US show conversion efficiencies from solar radiation to renewable energy above 20% (Miller and Keith, 2018). In other words, current solar energy technology is about a factor of 100 more efficient in converting solar radiation into renewable energy than wind or biomass. This high conversion efficiency is achieved because photovoltaics uses the low entropy of sunlight directly without the intermediate conversion into heat as in the case of atmospheric motion, which produces a lot of entropy without yielding free energy (Kleidon et al. 2016). Solar-based renewable energy technology, such as photovoltaics, thus has a much greater potential than other renewable forms of energy including wind energy because it is able to utilize the low entropy contained in solar radiation.

We can draw two kinds of conclusions from this analysis, on the theoretical as well as the applied side. On the theoretical side, we can conclude that a thermodynamic Earth system perspective, which follows the energy from the source to process, accounts for interactions, and uses thermodynamics to derive limits, can provide valuable insights about the main, first-order controls of the strength of the atmospheric circulation. Furthermore, it shows how the line of research over the last decades that used the proposed principle of Maximum Entropy Production may not be so far-fetched as it may seem, as it can also be interpreted as a manifestation of maximizing power. Whether it is, in the end, entropy production or power that is maximized is a relatively minor distinction that does not affect these conclusions. This approach should allow for further applications in the future, in which climatological questions can be addressed analytically to derive first-order explanations and estimates, complementing the insights gained from complex numerical simulation models.

On the applied side, this paper shows that such theoretical considerations as to how and how much the atmosphere generates kinetic energy has practical implications for wind energy yields and resource estimates. It explains how the atmospheric effects of using wind energy diminish the efficiency of wind turbines when wind energy use expands in scale. These declines in yields are likely to play an important role in the coming transition to a renewable energy system. A recent study by Agora Energiewende et al. (2020) showed how accounting for these effects, the expected yields of offshore wind energy may get diminished by a third or more in realistic energy scenarios for the year 2050 for Germany. To account for these effects, it does not necessarily require a complicated approach, but it does require a more inclusive, system-based approach that budgets the kinetic energy. This should lead to better, more realistic wind resource estimates that show that there is still a lot of renewable energy to be gained, but that detrimental atmospheric effects need to be accounted for.

# Appendix A

The limit to how much energy an individual wind turbine can take out of the atmospheric flow is well known as the Lanchester-Betz limit, named after two scientists active at the beginning of the 20th century who discovered this limit independently (Lanchester, 1915; Betz, 1920; see also van Kuik 2007). This limit follows from physical constraints, specifically, mass and momentum balances, two physical properties that are subjected to conservation laws. These balances set hard limits to how much kinetic energy a single wind turbine can at best take out of the atmosphere, irrespective of its technology.

To derive this limit, let us consider a setup as shown in Figure A1. The undisturbed atmospheric flow upwind of the turbine is described by a certain wind speed, $v_0$. When it encounters a wind turbine, the streamlines of the flow spread out - as shown by the near-horizontal lines in Figure A1, resulting in the slowing of the winds at the turbine, with a wind speed $v_t$, and in the area behind the rotor blades, with a wind speed of $v_w$ in the so-called wake. This reduction in wind speed is associated with the formation of a pressure difference, following Bernoulli's law in fluid dynamics. This pressure difference is what forces the rotor blades to turn, drive the generator of the turbine, and ultimately removes the kinetic energy from the flow.

The air flow is subjected to continuity, or mass conservation, so that no air mass is gained or lost as it flows through the cross-sectional area spanned by the rotor blades. In other words, when we follow this air mass as it flows through the rotor-swept area of the turbine, we can figure out how much kinetic energy was removed from the flow. The mass flow, $J_m$, is described by the density of the air $\rho$, the rotor-swept area of the turbine $A$, as well as the wind speed at the turbine, $v_t$, resulting in

$$J_m = \rho A v_t \tag{A1}$$

The wind speed at the turbine is given by the average of the undisturbed wind speed $v_0$ and the wind speed in the wake, $v_w$ (which can be derived in fluid dynamics by considering a so-called actuator disk, which was introduced by Froude). With this, we can express the mass flow as

$$J_m = \frac{1}{2} \rho A (v_0 + v_w) \tag{A2}$$

The kinetic energy that the turbine removed from the flow is obtained by considering the flow of kinetic energy associated with the mass flow before and behind the turbine. By taking the difference, we know how much kinetic energy was taken out of the flow, and how much power, $G_{turb}$, the turbine generated

$$G_{turb} = \frac{1}{2} \cdot J_m \cdot \left(v_o^2 - v_w^2\right) \tag{A3}$$

We can now combine Eq. (A2) and Eq. (A3) and express the reduction in wind speed in the wake by a ratio, $r = v_w/v_0$, which yields

$$G_{turb} = \frac{1}{2} \rho v_0^3 A \cdot \frac{1+r}{2} \cdot (1-r^2) \tag{A4}$$

The first term on the right-hand side is the flux of kinetic energy through the rotor-swept area, expressed in terms of the undisturbed wind speed $v_0$, while the other terms that involve $r$ express how much this flux is reduced by the turbine. A stronger reduction of the flow by the turbine (corresponding to a smaller value of $r$) reduces the second term on the right-hand side (representing the mass flow), but increases the third term (representing the kinetic energy). Less reduction has the reverse effect. Hence, this equation has a maximum in power $G_{turb}$ with respect to $r$.

Mathematically, this maximum in power is obtained by taking the derivative, $dG_{turb}/dr = 0$. It yields a solution of $r = -1$ (which is unphysical and can be disregarded) or $r = 1/3$. The latter solution then yields a maximum in power of

$$G_{turb} = \frac{1}{2} \rho v_0^3 A \cdot \frac{16}{27} \tag{A5}$$

This fraction of $16/27 \approx 59\%$ is known as the Lanchester-Betz limit of an individual wind turbine. It is associated with a reduction of the wind speed in the wake to

$$v_w = \frac{v_0}{3} \tag{A6}$$

so that the kinetic energy in the air flow behind the wind turbine has been reduced to 1/9th of its original value.

## A2. Dissipative loss by wake turbulence

What happens to the wake after the wind turbine extracted the kinetic energy from the flow? Eventually, the shear between the wake and the mean flow drives turbulent mixing towards re-establishing a flow with a uniform wind speed. This wind speed is referred to here as $v_\infty$. It should be somewhat less than $v_0$ because some kinetic energy was removed, unless kinetic energy was replenished from elsewhere, e.g., higher atmospheric layers.

During this mixing, kinetic energy from the mean flow is further dissipated by wake turbulence. It can be easily derived by extending the Betz limit, as shown by Corten (2001). During the mixing within the wake, momentum is conserved, while kinetic energy is dissipated. What we do in the following is to formulate the momentum balance first and then calculate how much energy was lost due to the turbulent mixing.

The total momentum of the air flow within the wake and the surrounding flow is conserved when the wind speeds mix to yield the uniform wind speed $v_\infty$. This can be described by the momentum balance,

$$\epsilon \cdot J_{m,0} \cdot v_w + (1-\epsilon) \cdot J_{m,0} \cdot v_0 = J_{m,0} \cdot v_\infty \tag{A7}$$

which consists of the momentum fluxes before (left hand side) and after (right hand side) the wake mixes. The parameter ε in Eq. 7 is the fraction that the wake flow occupies compared to the total cross section of the air flow, and $J_{m,0}$ is the mass flow upwind of the turbine. As we will see below, the actual value of $\varepsilon$, in approximation, does not matter. This equation yields an expression for $v_\infty$:

$$v_\infty = v_0 \cdot (1 - \epsilon(1-r)) \tag{A8}$$

where $r = v_w/v_0$, as above.

The dissipative loss, $D_{wake}$, due to the mixing is given by the difference in the kinetic energy flow before and after the mixing occurred. This is described by

$$D_{wake} = \epsilon \cdot \frac{1}{2} J_{m,0} \cdot v_w^2 + (1-\epsilon) \cdot \frac{1}{2} J_{m,0} \cdot v_0^2 - \frac{1}{2} J_{m,0} \cdot v_\infty^2 \tag{A9}$$

Using $v_w = r\, v_0$ and Eq. A7, we obtain

$$D_{wake} = \frac{1}{2} J_{m,0} \cdot \left(\epsilon \cdot r^2 v_0^2 + (1-\epsilon) \cdot v_0^2 - (1-\epsilon(1-r))^2 v_0^2\right)$$

or

$$D_{wake} \approx \epsilon \cdot \frac{1}{2} J_{m,0} \cdot (1-r)^2 \tag{A10}$$

To relate this dissipation rate back to how much the turbine generates we need to relate the mass flux of the undisturbed air flow, $J_{m,0}$, to the mass flux through the rotor-swept area of the turbine, $J_m$. This is done by noting that

$$\epsilon \cdot J_{m,0} = \rho\, v_t\, A = \rho\, v_0\, A \cdot \frac{1+r}{2} \qquad (A11)$$

Combined with Eq. (A10), this yields for the dissipative wake loss

$$D_{wake} \approx \frac{1}{2} \rho\, v_0^3\, A \cdot (1-r)^2 \cdot \frac{1+r}{2} \qquad (A12)$$

or, when set into proportion of the generated power by the turbine (Eq. A4)

$$\frac{D_{wake}}{G_{turb}} = \frac{1-r}{1+r} \qquad (A13)$$

At the Lanchester-Betz limit ($r = 1/3$), this corresponds to a wake loss of $D_{wake}/G_{turb} = 0.5$, that is, half of the generated energy by the turbine.

# Acknowledgements

The author thanks Kerry Emanuel and one anonymous reviewer for their constructive reviews, the NASA CERES team for making the satellite data openly available (doi: 10.5067/Terra-Aqua/CERES/EBAF_L3B.004.1) and the Copernicus Climate Change Service for the access to the ERA-5 reanalysis data (doi: 10.24381/cds.f17050d7).

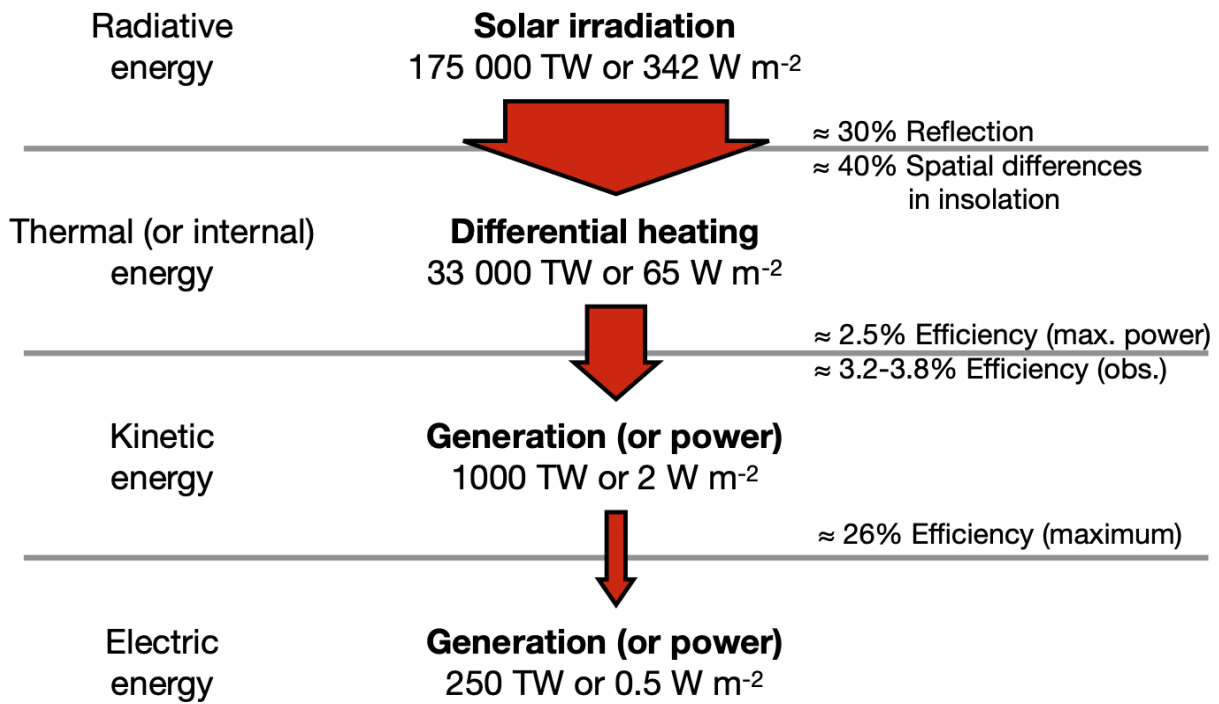

**Figure 1:** Schematic diagram of the different energy forms involved when solar radiation is converted to wind and renewable energy. Numbers provided are derived in the text.

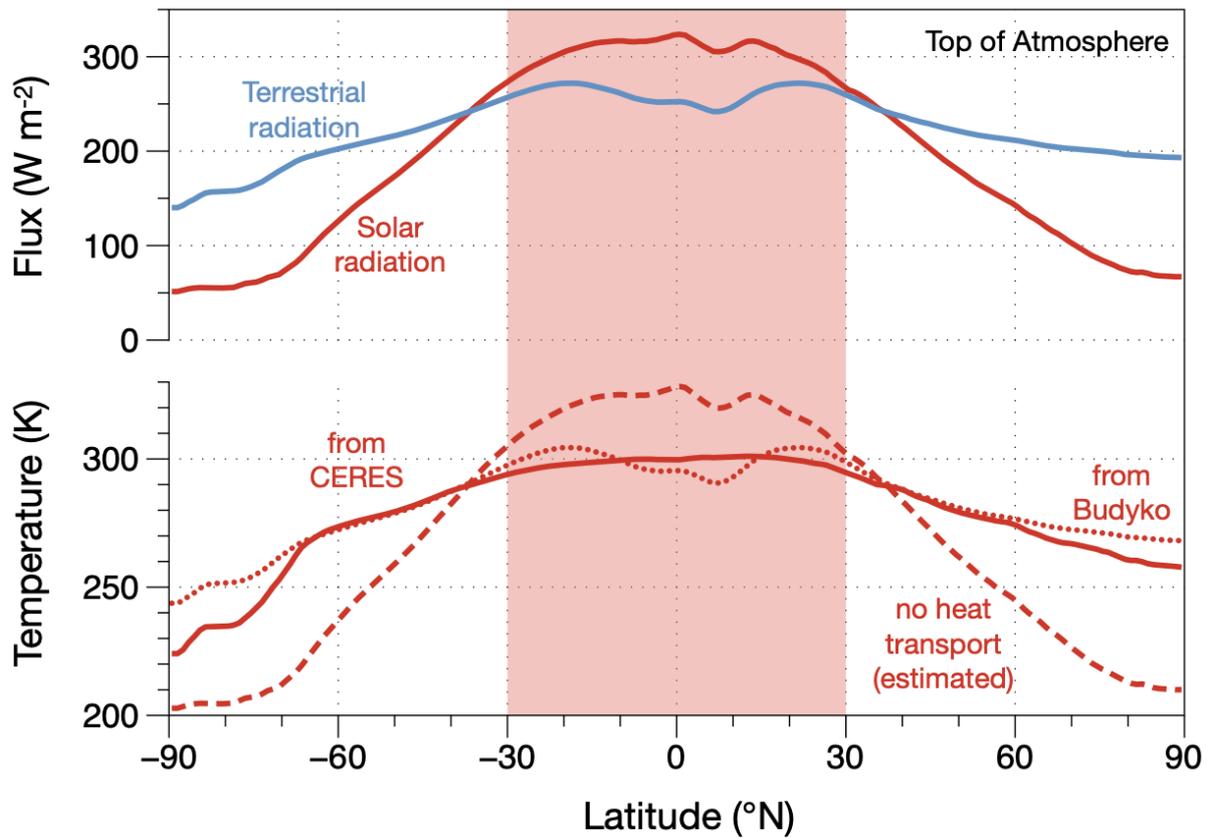

**Figure 2:** Zonal climatological means of the net fluxes of solar (red) and terrestrial (blue) radiation at the top of the atmosphere (top), taken from CERES satellite-derived estimates. The lower panel shows the associated surface temperatures: inferred from mean surface emission from the CERES estimates (solid line), inferred from Budyko's (1969) empirical parameterization using the TOA flux of terrestrial radiation (dotted line), and estimated for the case of no heat transport (dashed line). The red-shaded area marks the area that is referred to as "tropical" in the text, representing one half of the surface area of the Earth. The CERES data is available at doi:10.5067/TERRA-AQUA/CERES/EBAF-TOA_L3B004.1 and is described in Loeb et al. (2018) and Kato et al. (2018).

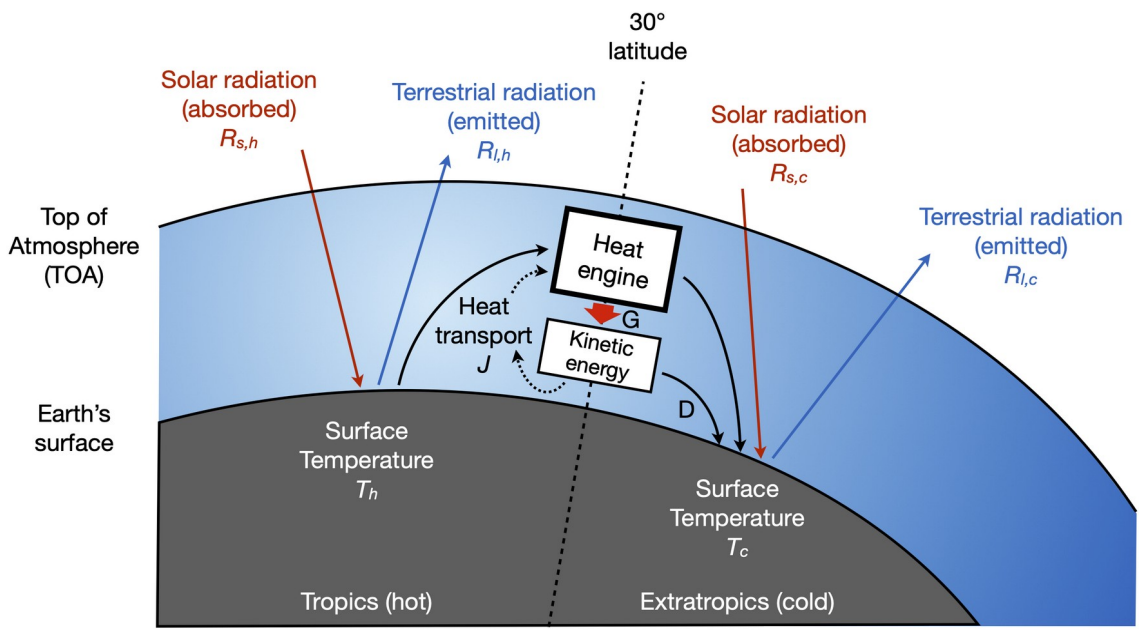

**Figure 3:** The atmospheric heat engine generates the kinetic energy (wide red arrow, with label *G*) associated with large-scale motion from the heat it transports from the tropics to the extratropics to balance out differences in absorbed solar radiation. The generated kinetic energy sustains the atmospheric motion that accomplishes the heat transport, as indicated by the dotted arrows. The different symbols next to the terms are used for the estimate described in the text.

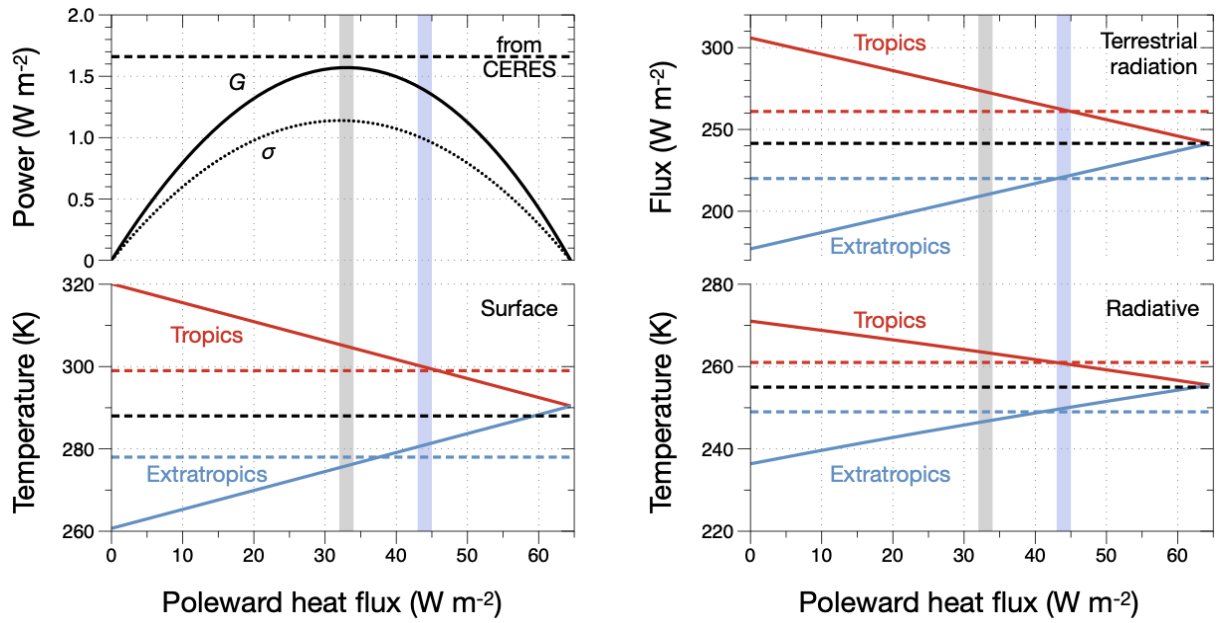

**Figure 4:** Sensitivity of kinetic energy generation of the two-box model described in the text to the magnitude of poleward heat transport. The four panels show power $G$ (solid line, Eq. 2.7, top left) and entropy production $\sigma$ (dotted line, Eq. 2.12, in units of 5 x mW m$^{-2}$ K$^{-1}$, top left), outgoing longwave radiation at the top of the atmosphere, $R_{l,h}$ and $R_{l,c}$, (top right), surface temperatures $T_h$ and $T_c$ (Eqs. 2.3 and 2.4, bottom left) and radiative temperatures (bottom right). Also shown are the values associated with maximum power (grey shaded area) and values derived from the CERES dataset: poleward heat transport (vertical blue shaded area), power inferred from heat flux and temperature difference (black dashed line), and radiative flux and temperatures (horizontal dashed lines).

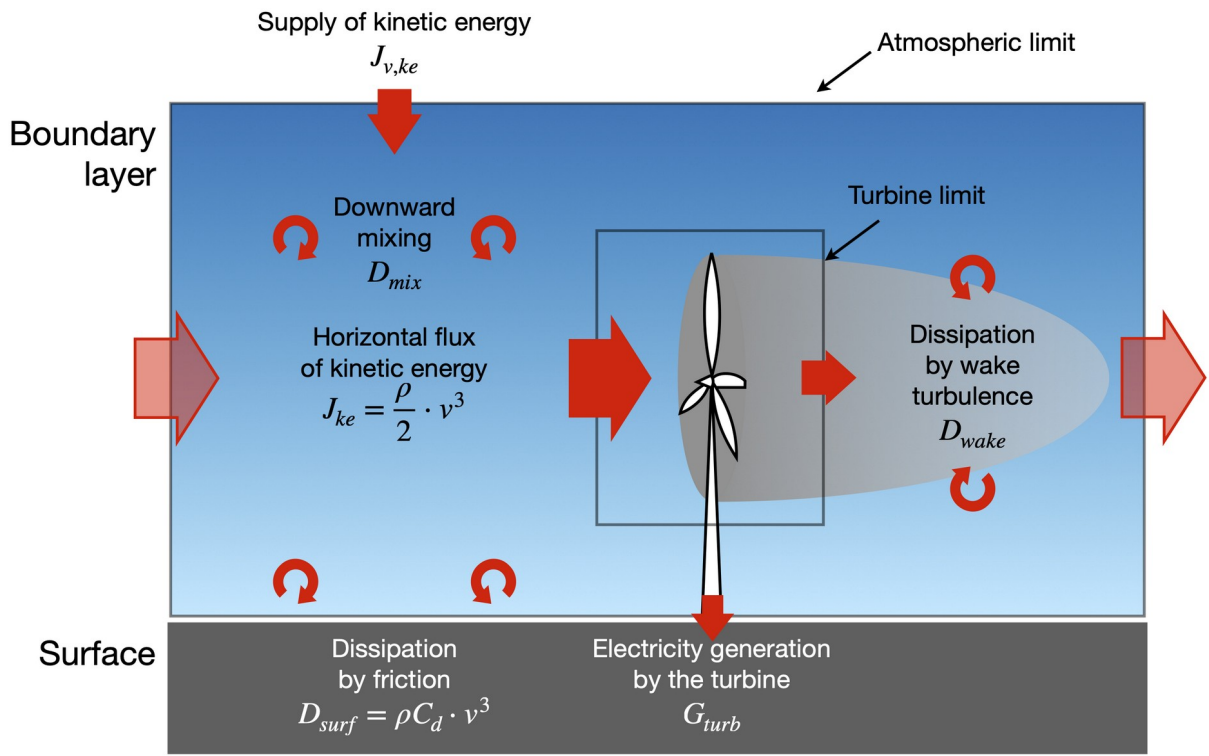

**Figure 5:** Schematic diagram to illustrate kinetic energy transport, dissipation, and use by wind turbines in the atmosphere near the Earth's surface.

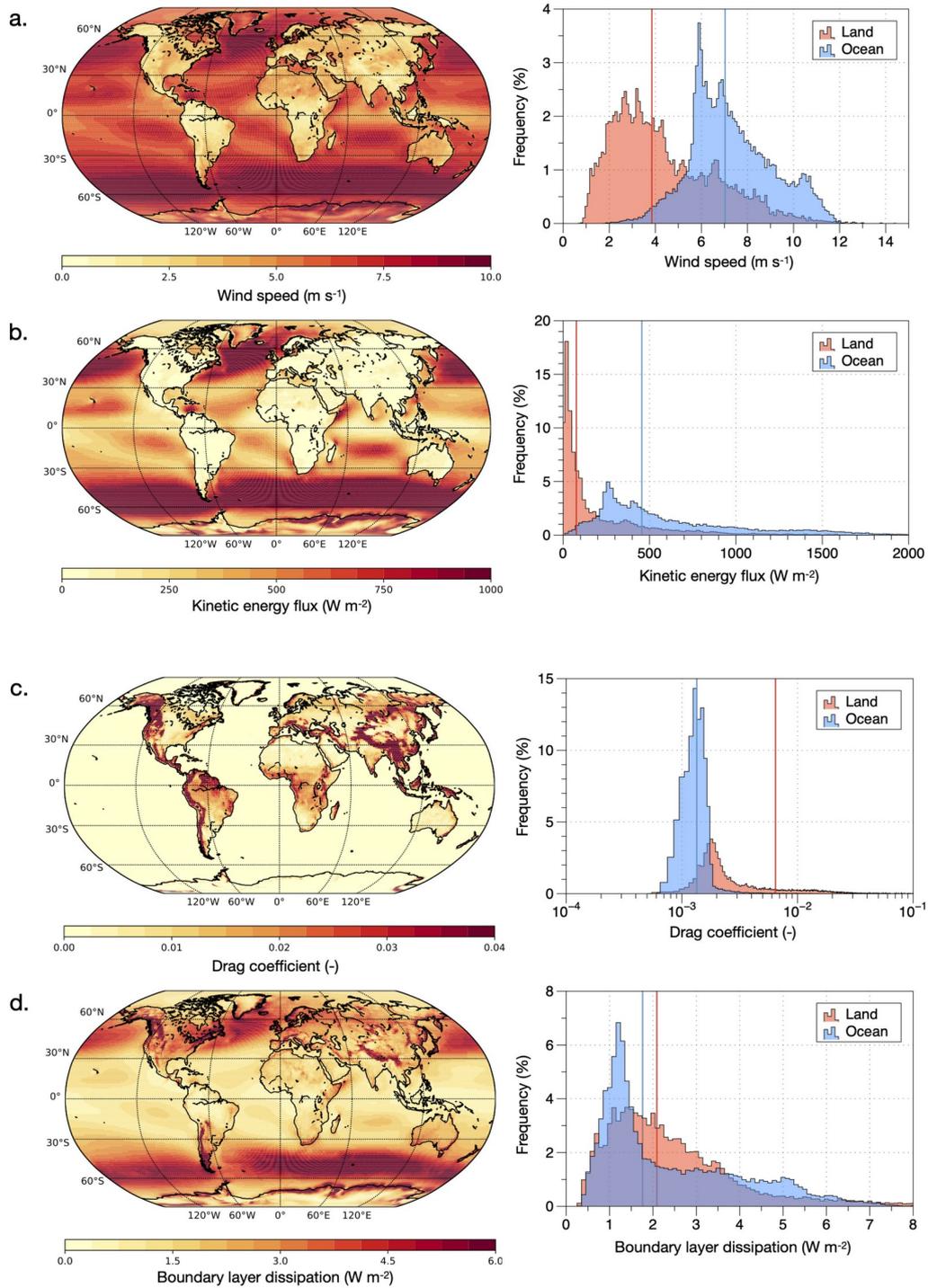

**Figure 6:** Climatological mean fields (1980-2009) of the ERA5-Reanalysis of (a) 10m wind speed, (b) kinetic energy flux density, (c) drag coefficient, and (d) boundary layer dissipation as well as their histograms for land (red) and ocean (blue) grid cells. The median values are marked by the vertical lines in red (land) and blue (ocean). Generated using Copernicus Climate Change Service (2017) information.

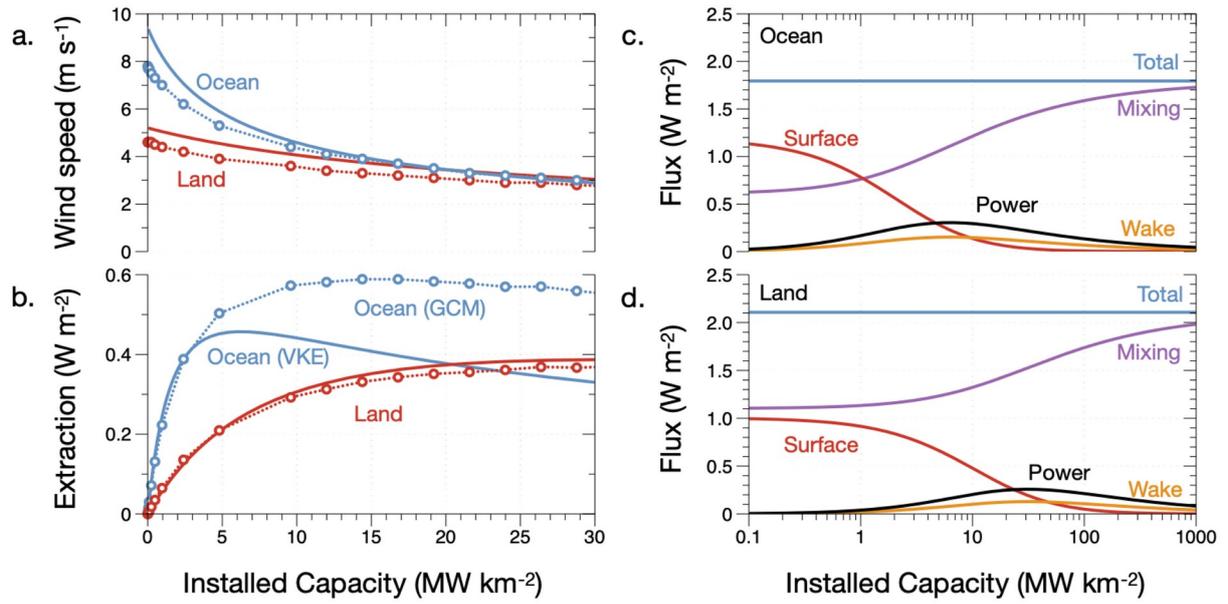

**Figure 7:** a.) Sensitivity of wind speeds and b.) extracted kinetic energy by wind turbines ($G_{turb} + D_{wake}$) at different installed capacities when wind energy is used at the planetary scale. The dotted lines show the sensitivities estimated from global climate model simulations by Miller and Kleidon (2016), with each circle representing one sensitivity simulation. The solid lines represent the sensitivity estimated by the analytical model described in the text ("VKE", Miller et al., 2011, Miller et al., 2015). The plots on the right show the associated kinetic energy budgets for (c.) ocean and (d.) land, with the fluxes $J_{v,ke}$ ("Total", blue), $D_{mix}$ ("Mixing", purple), $D_{surf}$ ("Surface", red), $D_{wake}$ ("Wake", orange), and the power generated by wind turbines, $G_{turb}$ ("Power", black), as described in Section 3.1. The lines show climatological means averaged over land and ocean separately.

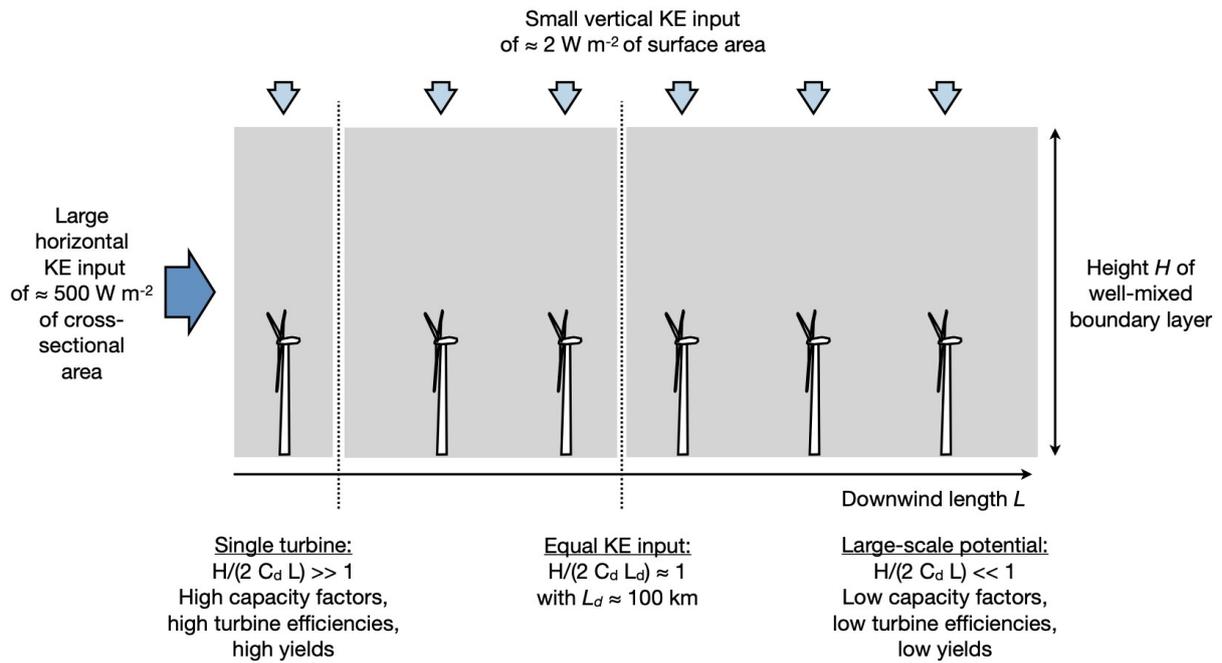

**Figure 8:** Illustration of the declining stock of kinetic energy in the lower atmosphere as wind energy use increases in scale, resulting in lower turbine efficiencies, yields, and wind energy resource potentials. The blue arrows show qualitatively the kinetic energy supplied by the horizontal flow (dark blue) and the vertical replenishment (light blue).

| Wind Energy | Bioenergy | Solar Energy |
|---|---|---|
| 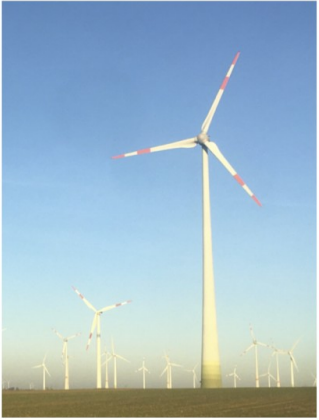 | 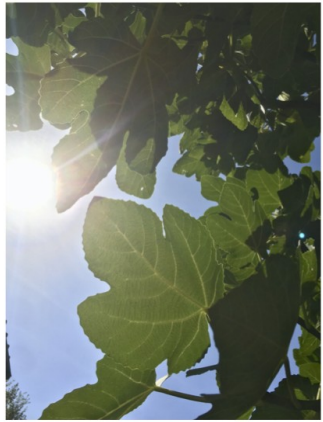 | 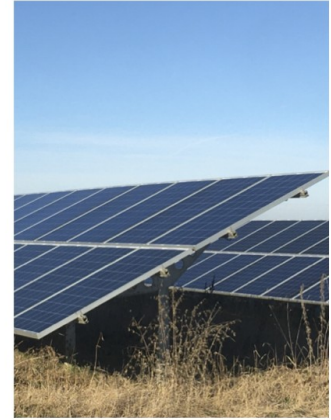 |
| <u>Total efficiency</u><br>Solar ➔ Wind: 0.7%<br>Wind ➔ Electricity: < 26%<br>Total: 0.2% | <u>Total efficiency</u><br>Solar ➔ Biomass: 0.4%<br>Biomass ➔ Electricity: < 60%<br>Total: 0.2% | <u>Total efficiency</u><br>Solar ➔ Electricity: 20%<br>Total: 20% and more |

**Figure 9:** Comparing wind energy to bioenergy and solar energy in terms of their total conversion efficiency from the source of solar radiation to renewable energy in form of electricity.

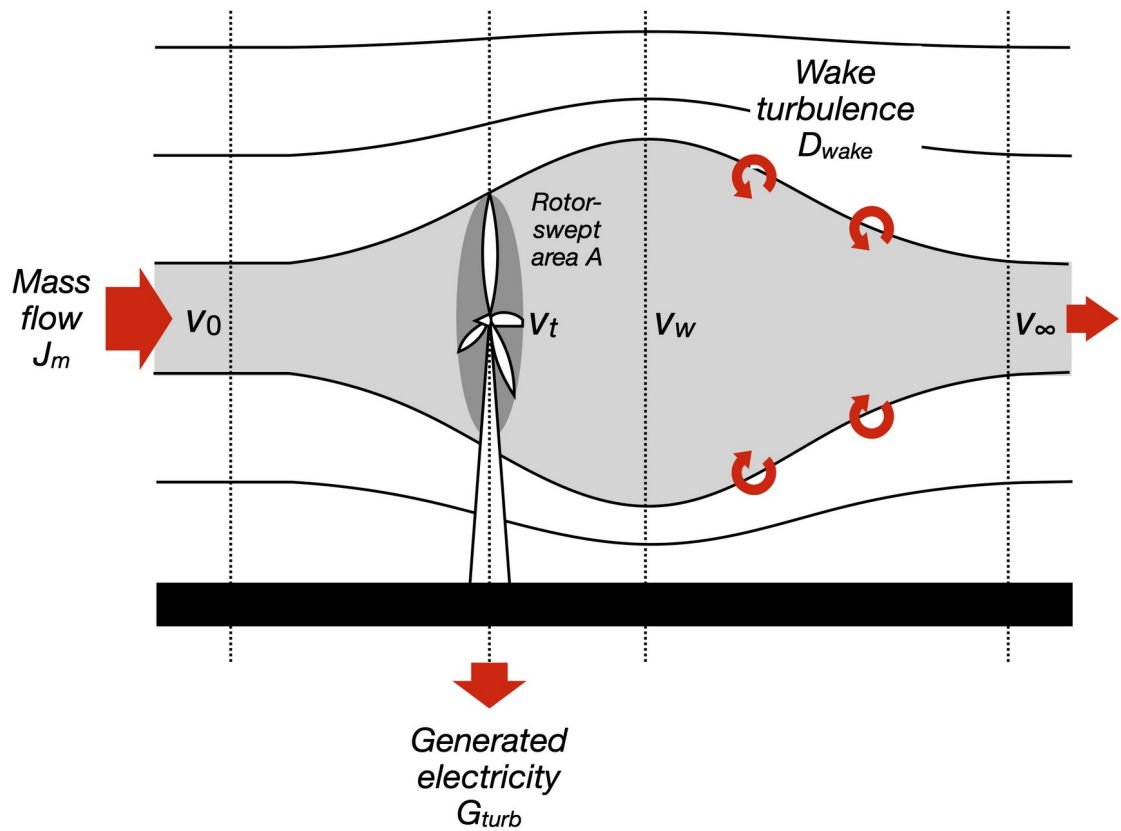

**Figure A1:** Schematic diagram of a wind turbine and the surrounding mass flow, with different wind velocities of the unaffected flow ($v_0$), at the turbine ($v_t$), within the wake ($v_w$) and after the wake was dissolved by wake turbulence ($v_\infty$).

# Tables

**Table 1:** Symbols used to estimate the maximum generation of kinetic energy associated with the large-scale atmospheric circulation as well as their description and numerical values.

| Symbol | Description | Value | Source |
|---|---|---|---|
| $R_{s,h}$ | Mean absorption of solar radiation in the tropics | 306 W m$^{-2}$ | from CERES climatology |
| $R_{s,c}$ | Mean absorption of solar radiation in the extratropics | 177 W m$^{-2}$ | from CERES climatology |
| $a$ | Empirical constant for longwave parameterization | -388.7 W m$^{-2}$ | Budyko (1969) |
| $b$ | Empirical constant for longwave parameterization | 2.17 W m$^{-2}$ K$^{-1}$ | Budyko (1969) |
| $J_{opt}$ | Mean poleward heat transport derived from max. power | 32 W m$^{-2}$ or 16.5 PW | Eq. (2.8) |
| $J_{CERES}$ | Mean poleward heat transport derived from CERES | 44 W m$^{-2}$ or 22.5 PW | from CERES climatology |
| $T_{h,opt} - T_{c,opt}$ | Mean temperature difference derived from max. power | 30 K | Eq. (2.9) |
| $T_h - T_c$ | Mean temperature difference derived from CERES | 21 K | from CERES climatology |
| $G_{max}$ | Generation of kinetic energy derived from max. power | 1.6 W m$^{-2}$ or 800 TW | Eq. (2.10) |
| $G_{CERES}$ | Generation of kinetic energy derived from CERES | 1.7 W m$^{-2}$ or 850 TW | from CERES climatology |
| $\eta$ | Efficiency in generating kinetic energy (power divided by poleward heat transport) | 5 % | Eq. (2.11) |

**Table 2:** Symbols used to estimate the maximum generation of renewable energy from wind at large scales as well as their description and numerical values.

| Symbol | Description | Ocean | Land | Source |
|---|---|---|---|---|
| $v$ | Wind speed at 10m height | 7.0 m s$^{-1}$ | 3.9 m s$^{-1}$ | from ERA-5 (Fig. 6a) |
|  |  | 9.4 m s$^{-1}$ | 5.2 m s$^{-1}$ | from $J_{ke}$ (Fig. 6b) |
| $J_{ke}$ | Horizontal kinetic energy flux density (per cross-sectional area) | 455 W m$^{-2}$ | 77 W m$^{-2}$ | from ERA-5 (Fig. 6b) |
| $\rho$ | Air density | 1.1 kg m$^{-3}$ | 1.1 kg m$^{-3}$ |  |
| $C_d$ | Drag coefficient | 0.0013 | 0.0065 | from ERA-5 (Fig. 6c) |
| $F_{down}$ | Downward transport of horizontal momentum | 0.13 N m$^{-2}$ | 0.19 N m$^{-2}$ | from $\tau$ with no wind turbines |
| $J_{v,ke}$ | Downward kinetic energy flux density (per surface area) | 1.8 W m$^{-2}$ | 2.1 W m$^{-2}$ | from ERA-5 (Fig. 6d) |
| $v_{free}$ | Wind speed of the free atmosphere | 14.2 m s$^{-1}$ | 10.9 m s$^{-1}$ | inferred from $J_{v,ke}$ and surface drag, Eq. 3.6 |
| $F_{turb,opt}$ | Optimum turbine drag to maximize yield | 0.08 N m$^{-2}$ | 0.13 N m$^{-2}$ | $G_{turb,max}/v_{opt}$ |
| $v_{opt}$ | Wind speed at maximum yield | 5.4 m s$^{-1}$ | 3.0 m s$^{-1}$ | Eq. 3.12 |
| $G_{turb,max}$ | Maximum yield/electricity generation (per surface area) | 0.30 W m$^{-2}$ | 0.26 W m$^{-2}$ | Eq. 3.13 |
| $D_{wake}$ | Wake dissipation (per surface area) | 0.15 W m$^{-2}$ | 0.13 W m$^{-2}$ | 1/2 $G_{turb}$ as in Lanchester-Betz limit (see Eq. A13) |
| $D_{surf}$ | Dissipation by surface friction (per surface area) | 0.23 W m$^{-2}$ | 0.19 W m$^{-2}$ | Eq. 3.8 |
| $D_{mix}$ | Dissipation by mixing above the turbines (per surface area) | 1.14 W m$^{-2}$ | 1.5 W m$^{-2}$ | Eq. 3.7 |
| $\eta_{wind}$ | Efficiency of wind energy use | 16.7% | 12.4% | $G_{turb,max}/J_{v,ke}$ |